\documentclass[journal,doublecolumn]{IEEEtran}

\usepackage{indentfirst}
\usepackage[pdftex]{graphicx}
\usepackage{amsthm}
\usepackage[fleqn]{amsmath}
\usepackage{amssymb}
\usepackage{amsfonts}
\usepackage{mathrsfs}
\usepackage{graphicx}
\usepackage{enumerate}
\usepackage{multirow}
\usepackage{enumitem}
\usepackage{enumerate}   
\usepackage{float}  
\usepackage{tabularx} 
\usepackage{threeparttable}
\usepackage{booktabs} 
\usepackage{makecell} 
\usepackage[noend]{algpseudocode}
\usepackage{algorithmicx,algorithm}
\usepackage{bm}

\newtheorem{Theorem}{Theorem}
\newtheorem{Definition}{Definition}
\newtheorem{Proposition}{Proposition}
\newtheorem{Corollary}{Corollary}

\newtheorem{Remark}{Remark}

\newtheorem*{Square Root Law}{Square Root Law}
\newtheorem{Lemma}{Lemma}

\newtheorem*{Proof of Theorem 2}{Proof of Theorem 2}

\ifCLASSINFOpdf
\else
\fi

\begin{document}
\setlength{\abovedisplayskip}{6pt}
\setlength{\belowdisplayskip}{4pt}
%
\title{First- and Second-Order Bounds for Covert Communication Over AWGN Channels}
%
%
%

\author{Xinchun~Yu, Chenhao Ying,
  Shuangqing Wei and Xiao-Ping Zhang 
\thanks{Xinchun Yu is with Department of Computer Science and Technology, Zhejiang GongShang University (e-mail: yxc@mail.zjgsu.edu.cn). Chenhao Ying is with Department of Computer Science, Shanghai Jiao Tong University, Shanghai
200240, China (e-mail: yingchenhao@sjtu.edu.cn). Shuangqing Wei is with the Division of Electrical and Computer Engineering
School of Electrical Engineering and Computer Science,
Louisiana State University, Baton Rouge, LA 70803, USA (e-mail: swei@lsu.edu). Xiao-Ping Zhang is with Institute of Information and Data Science, Shenzhen International
Graduate School, Tsinghua University, Shenzhen, China (e-mail: xpzhang@ieee.org). Corresponding author: Xiao-Ping Zhang. 
 
}
}

\maketitle



%
\IEEEpeerreviewmaketitle

\begin{abstract}
This paper investigates the first- and second-order asymptotic behavior of covert communication over additive white Gaussian noise (AWGN) channels under a Kullback--Leibler (KL) divergence covertness constraint. We derive explicit achievable first- and second-order throughput terms and a converse bound in terms of the blocklength, decoding error probability, adversary noise variance, and covertness constraint. The achievability analysis employs truncated Gaussian random coding with an arbitrary fixed parameter $\mu\in(1/2,1)$, a common decoding threshold accommodating codewords with different energies, and expurgation to meet a maximal-error reliability requirement. We also establish sufficient secret-key sizes and identify the dependence of key consumption on the adversary noise variance and $\mu$. Both throughput bounds exhibit $O(\sqrt{n})$ first-order and $O(n^{1/4})$ second-order scalings. Although their coefficients differ for each fixed $\mu<1$, the ratio of the achievable to converse first-order coefficients equals $\mu$. Thus, choosing a fixed $\mu$ arbitrarily close to one makes the first-order coefficient gap arbitrarily small. The results quantify the tradeoffs among throughput, reliability, covertness, and secret-key resources in finite-blocklength AWGN covert communication.
\end{abstract}


\section{Introduction}
Covert communication, also known as communication with low probability of detection (LPD), aims to conceal the presence of legitimate transmissions from an adversary. Unlike conventional secure communication, which protects message contents, covert communication seeks to prevent an adversary from detecting communication activities. It has attracted considerable attention in wireless communications, particularly in privacy-sensitive and adversarial environments \cite{Matthieu R,Bloch3,Jaggi,Chen,Yan2}.

The fundamental limits of covert communication were initially investigated over additive white Gaussian noise (AWGN) channels \cite{Boulat A} and discrete memoryless channels (DMCs) \cite{Matthieu R,Ligong Wang}. A central result is the square-root law, which establishes that the maximal covert throughput scales as $O(\sqrt{n})$ bits over $n$ channel uses under standard channel and covertness assumptions. Subsequent studies extended covert communication to binary symmetric channels \cite{Pak Hou Che}, MIMO AWGN channels \cite{Abdelaziz,Abdelaziz2,Shi-Yuan Wang}, non-coherent fading channels \cite{M.Tahmasbi3}, Poisson channels \cite{Ligong Wang2}, and infinite-bandwidth Gaussian channels \cite{Ligong3}.

Beyond point-to-point channels, covert communication has been investigated in various network settings, including multiple-access channels \cite{Arumugam2019}, broadcast channels \cite{Arumugam2019_BC,Kibloff2019,Tan2019}, interference channels \cite{Cho2021_TIFS}, and adversarially jammed channels \cite{Jaggi}. Recent investigations have further examined covert capacity--key tradeoffs in DMC networks \cite{Bounhar} and the benefits of non-covert users in facilitating covert communication \cite{Bounhar2024}. Related research has also extended covert communication to quantum systems, including covert quantum sensing \cite{M.Tahmasbi4}, bosonic channels \cite{Gagatsos2020}, and entanglement-assisted covert communication \cite{Zlotnick2025,Wang2024}. Entanglement-assisted transceiver design has been investigated in \cite{Cox2023}. These developments demonstrate the broad applicability of covert communication and the importance of understanding its fundamental performance limits.

Although the square-root law describes the leading-order throughput scaling, it does not fully capture the effects of finite blocklength, decoding reliability, and covertness requirements. Finite-blocklength information theory \cite{Polyanskiy} provides a framework for quantifying these effects through higher-order asymptotic analysis. Tahmasbi and Bloch \cite{M.Tahmasbi,M.Tahmasbi2} investigated the first- and second-order asymptotics of covert communication over DMCs using channel resolvability, considering KL divergence and other covertness metrics.

For AWGN channels, our previous works \cite{Yu0,Yu1} developed one-shot achievability and converse bounds for Gaussian random coding under power constraints. Related second-order asymptotic results were also investigated in \cite{Yuicc}. Under a KL-divergence covertness constraint, obtaining explicit achievable throughput coefficients together with sufficient secret-key requirements for truncated Gaussian coding calls for a joint treatment of reliability and resolvability. In particular, truncation produces codewords with different energies, so a common decoding threshold must control their error probabilities; meanwhile, the resolvability analysis must account for the truncation normalization and the adversary's noise variance. The subsequent expurgation step must preserve both maximal-error reliability and covertness. These issues motivate the analysis developed here.

In this paper, we study covert communication over AWGN channels under maximal decoding error probability $\epsilon$ and KL-divergence covertness constraint $\delta$. We establish explicit achievable first- and second-order throughput terms, a converse bound, and sufficient secret-key sizes. A central feature of the achievability result is that the truncation parameter $\mu$ may be selected as \emph{any fixed constant} in $(1/2,1)$. The achievable and converse first-order coefficients are, respectively, $\mu\sigma^2\sqrt{\delta\log e}$ and $\sigma^2\sqrt{\delta\log e}$. Consequently, their ratio is exactly $\mu$: for every $\eta\in(0,1/2)$, a fixed choice $\mu\in(1-\eta,1)$ makes the achievable first-order coefficient exceed $(1-\eta)$ times the converse coefficient. Thus, the proposed family of coding constructions approaches the converse first-order coefficient arbitrarily closely, even though the two coefficients do not coincide for any admissible fixed $\mu$.

Our main contributions are summarized as follows.
\begin{enumerate}
\item \emph{Achievable throughput and converse bounds:} We derive explicit achievable first- and second-order throughput terms and a converse bound as functions of the blocklength $n$, decoding error probability $\epsilon$, adversary noise variance $\sigma^2$, and covertness constraint $\delta$. Both bounds exhibit $O(\sqrt{n})$ first-order and $O(n^{1/4})$ second-order scalings. Since $\mu$ is arbitrary and fixed in $(1/2,1)$, the relative gap between the first-order coefficients can be made arbitrarily small, although the coefficients are not equal for a given $\mu<1$.

\item \emph{Secret-key requirements:} For fixed $\mu\in(1/2,1)$ and $\epsilon\in(0,1/2)$, we establish sufficient secret-key lengths of order $\sqrt n$ when $\sigma^2<\mu^{-2}$ and order $n^{1/4}$ when $\sigma^2=\mu^{-2}$. When $\sigma^2>\mu^{-2}$, the construction requires no secret-key bits for all sufficiently large blocklengths. 

\item \emph{Reliability, resolvability, and expurgation analysis:} We develop an achievability framework based on independently generated truncated Gaussian codewords, threshold decoding, and expurgation. By exploiting spherical symmetry and selecting a common decoding threshold, we obtain uniform Type-I error bounds for codewords with varying energies. We also derive an ensemble-average KL-divergence bound through a direct resolvability analysis. Concentration and expurgation arguments then establish the existence of codebooks satisfying both maximal reliability and covertness constraints.
\end{enumerate}
 The preceding first-order approximation statement concerns a family of asymptotic results, each established with a fixed $\mu$. It does not imply that the remainder terms are uniform over $\mu$ near one or justify selecting a blocklength-dependent sequence $\mu_n\to1$ in the existing proof. Such a strengthening would require uniform control of reliability, resolvability, and expurgation estimates. Likewise, the present bounds do not establish matching second-order coefficients.

Compared with existing DMC results \cite{M.Tahmasbi2,Bounhar} and previous AWGN analyses \cite{Yu0,Yu1}, the present work explicitly accounts for Gaussian truncation in the achievability analysis and quantifies its effects on throughput and sufficient secret-key sizes. Our contribution is an explicit achievability analysis for truncated Gaussian coding under the stated power, reliability, and covertness constraints, together with sufficient secret-key scaling and a converse bound; we do not claim that all first- and second-order AWGN covert asymptotics are new. A comparison with related studies is provided in Table~\ref{tab:comparison}. Practical waveform and spectral constraints, as investigated in \cite{Qiaosheng1}, are beyond the scope of this work.

\begin{table*}[t]
    \centering
    \small
    \setlength{\tabcolsep}{5pt}
    \renewcommand{\arraystretch}{1.15}
    \begin{threeparttable}
        \caption{Comparison of achievability results with related works.}
        \label{tab:comparison}
        \begin{tabularx}{\textwidth}{@{}
            l
            >{\raggedright\arraybackslash}p{4.2cm}
            >{\raggedright\arraybackslash}X
            >{\raggedright\arraybackslash}X
            >{\raggedright\arraybackslash}X
        @{}}
            \toprule
            Reference & Channel / Noise assumptions & Achievable code-size asymptotics
            & Techniques & Sufficient key-size scaling \\
            \midrule
            \cite{Yu1}
            & AWGN channel; $\sigma_b^2 = \sigma_w^2 = 1$
            & No explicit first- or second-order asymptotics
            & Sequential random encoding; finite blocklength analysis
            & Not considered$^{\dagger}$ \\
            \midrule
            \cite{Yuicc}
            & AWGN channel; $\sigma_b^2 = \sigma_w^2 = 1$
            & First-order: $\sqrt{n\delta\log e}$;\newline
              Second-order: $\sqrt{2}(n\delta)^{1/4}(\log e)^{3/4}$
            & Coding techniques in \cite{Yu1}; local information geometry
            & Not considered$^{\dagger,\ast}$ \\
            \midrule
            \cite{Boulat A}
            & AWGN channel; $\sigma_b^2, \sigma_w^2$ known
            & First-order: $O(\sqrt{n})$
            & i.i.d.\ random coding; capacity analysis
            & First-order: $O(\sqrt{n}\log n)$ \\
            \midrule
            \cite{Ligong Wang}
            & AWGN channel; $\sigma_b^2 = \sigma_w^2 = \sigma^2$
            & First-order: $\sqrt{n\delta\log e}$
            & i.i.d.\ random coding; capacity analysis
            & Not considered$^{\dagger,\ddagger}$ \\
            \midrule
            This work
            & AWGN channel; $\sigma_b^2 = 1$, $\sigma_w^2 = \sigma^2$
            & $A_\mu\sqrt n
+B_\mu n^{1/4}\Phi^{-1}(\epsilon)
+o(n^{1/4})$
            & i.i.d.\ random coding; finite blocklength analysis
            &  \begingroup
\scriptsize
\(
\log K_n =
\begin{cases}
\bar A_\mu\sqrt{n}+o(\sqrt{n}),
  & \bar A_\mu>0,\\[2pt]
o(\sqrt{n}),
  & \bar A_\mu=0,\\[2pt]
0,
  & \bar A_\mu<0.
\end{cases}
\)
\tnote{e}
\endgroup \\
            \bottomrule
        \end{tabularx}
        \smallskip
        \begin{tablenotes}
            \footnotesize
            \item[$\dagger$] Secret-key size is not analyzed.
            \item[$\ddagger$] Only first-order asymptotics are treated; the truncation effect is not considered.
            \item[$\ast$] Second-order asymptotics are provided, but the truncation effect is not explicitly analyzed.
            \item[$\S$] $\sigma^2$ denotes the normalized noise variance; $\mu$ is defined in Sec.~\ref{preliminarydistributions}.
             \item[e] The key-size entry for this work
summarizes first-order sufficient scaling.
The sharper $n^{1/4}$ terms, including the
boundary case $\bar A_\mu=0$, are given
in Theorem~\ref{achievability0}.
The parameter $\mu\in(1/2,1)$ is fixed
independently of $n$.
        \end{tablenotes}
    \end{threeparttable}
\end{table*}

The rest of this paper is organized as follows. In Section II, we present some preliminaries on notations, the channel model, the outline of the proof of the achievability and the converse, the definitions of codes and some distributions. Section III presents the main results, which include the first and second order asymptotics of the achievability bound and the investigation of the converse bound. In Section IV, numerical results illustrate the effects of system parameters on throughput, the achievability–converse gap, and sufficient secret-key scaling. Section V concludes the paper. 
\section{Preliminaries}\label{model}
\subsection{Notations}
Throughout the paper, $\log$ denotes logarithmic function with base $2$. We use upper case letters, such as $X$, for random variables and lower case letters, such as $x$ for their realizations. Moreover, upper case letters with boldface fonts are used for random vectors, e.g., $\bm{Z}$, and their realizations are denoted with lower case letters of boldface font, e.g., $\bm{z}$. The norm $\|\cdot\|$ will be used for $l^2$ norm $\|\cdot\|_2$. We use $*$ for convolution operation and $\langle M \rangle$ for the set $\{1,\cdots,M\}$. If $a$ is a real number, then $|a|$ denotes its absolute value. If $\mathcal{X}$ is a set, then $|\mathcal{X}|$ denotes its cardinality. The function $\Phi$ denotes the cumulative distribution function of standard normal random variable.  For two distributions $P$ and $Q$ in the same probability space, some related divergences in this paper are defined as follows. 

The Kullback-Leibler divergence (KL divergence) 
\begin{equation}
\mathbb{D}(P\|Q) \triangleq \int \log \frac{dP}{dQ}dP.
\end{equation}
The total variational distance (TVD)
\begin{equation}
\mathbb{V}(P,Q) = \frac{1}{2} \int |\frac{dP}{dQ} -1|dQ.
\end{equation}

\subsection{The Channel Model}
\begin{figure}
\centering
\includegraphics[width=3.5in]{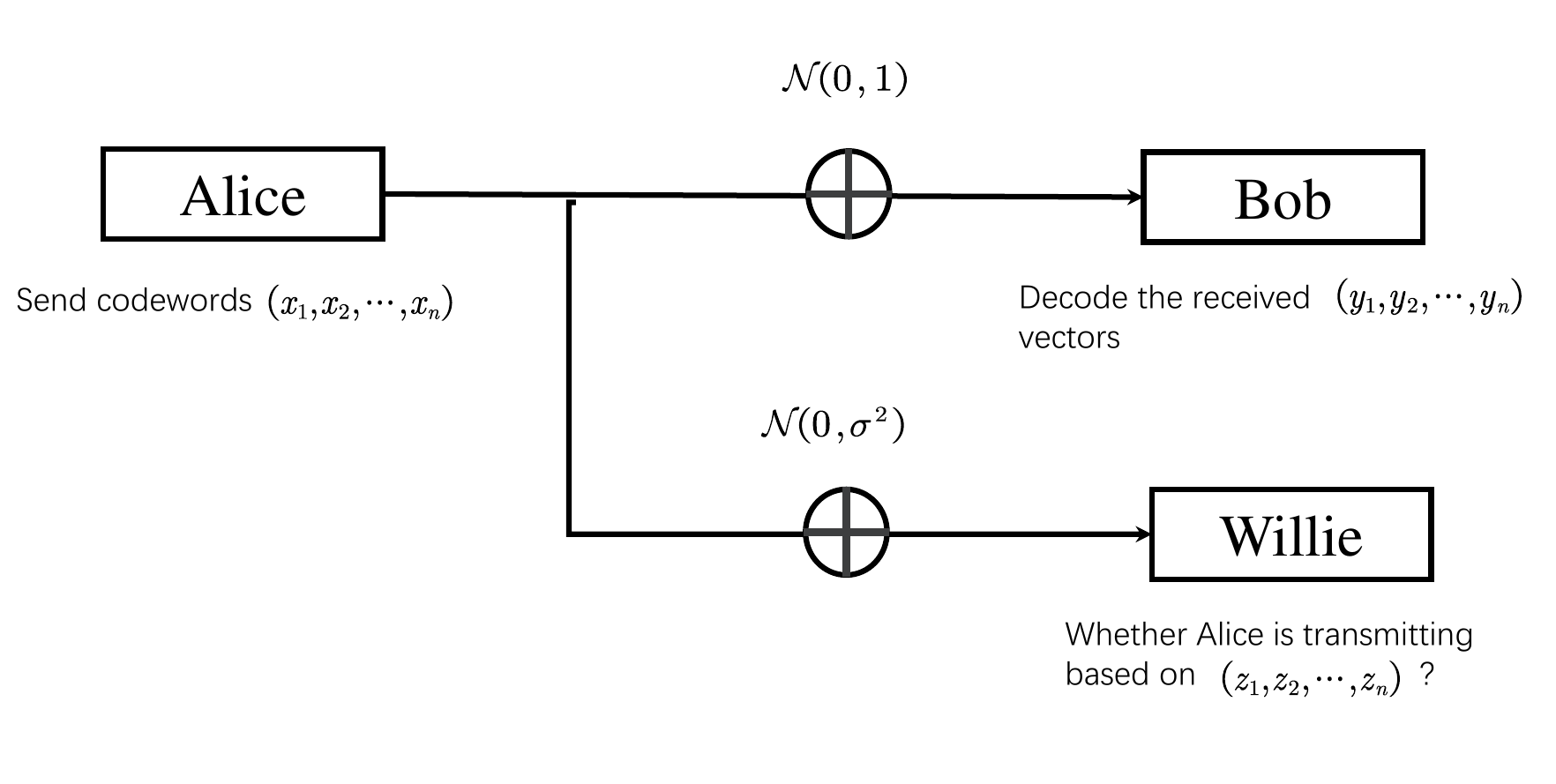}
\caption{The model of covert communication over AWGN channels in Section \ref{model}}\label{Fig1}
\end{figure}
The AWGN covert channel model considered in this work is illustrated in Fig.\ref{Fig1}. A transmitter Alice communicates with a legitimate receiver Bob through a main channel $W_1$, and the signals are also observed by an adversary Willie through another channel $W_2$, which are modeled as \footnote{The model also implies the noise powers at the main channel and the noise power at the adversary are $\sigma_b^2 = 1$ and $\sigma_w^2 = \sigma^2$, respectively.}
\begin{equation}\label{channelmodel0}
W_1(y|x) = \frac{1}{\sqrt{2\pi}} \exp \left(-\frac{(y-x)^2}{2}\right),
\end{equation}
\begin{equation}\label{channelmodel1}
W_2(z|x) = \frac{1}{\sqrt{2\pi\sigma^2}} \exp \left(-\frac{(z-x)^2}{2\sigma^2}\right).
\end{equation} 
The messages are indexed by $w \in \langle M \rangle $. 
A codeword over the AWGN channel is represented by a vector in $\mathbb{R}^n$: $\bm{x} = (x_1, x_2, \cdots, x_n)$. With channel transition models $W_1(y|x)$ and $W_2(z|x)$, we have 
$W^n_1(y^n|x^n) = \prod_{i=1}^n W_1(y_i|x_i)$ and $W^n_2(z^n|x^n) = \prod_{i=1}^n W_2(z_i|x_i)$.

\subsection{Definitions of Codes for AWGN Covert Channels}

We first define conventional AWGN codes and
then extend the definition to covert communication.

\begin{Definition}\label{definition1}
An $(M,n,\epsilon,\Psi)_m$ code for the AWGN
channel (\ref{channelmodel0}) consists of
an encoder
$f_n:\langle M\rangle\to\mathbb{R}^n$
and a decoder
$g_n:\mathbb{R}^n\to\langle M\rangle$
satisfying
\begin{equation}
P_{\rm err}\triangleq
\max_{w\in\langle M\rangle}
\mathbb{P}(\hat W\neq w\mid W=w)
\leq\epsilon
\end{equation}
and the maximal power constraint
\begin{equation}\label{maximalpower}
\|\bm{x}_w\|^2\leq n\Psi,
\quad w\in\langle M\rangle,
\end{equation}
where $\bm{x}_w=f_n(w)$.
Replacing (\ref{maximalpower}) with
\begin{equation}\label{averagepower}
\frac{1}{M}\sum_{w=1}^M
\|\bm{x}_w\|^2\leq n\Psi
\end{equation}
defines an $(M,n,\epsilon,\Psi)_a$ code.
\end{Definition}

Alice and Bob share a secret key $S$ selecting one of
$K$ publicly known subcodes of a deterministic code
$\bm c$. Willie knows the code but not the realized
key. The message $W$ and key $S$ are independent and
uniform over $\langle M\rangle$ and $\langle K\rangle$,
respectively. Covertness is imposed on Willie's output
distribution averaged over both indices.

\begin{Definition}\label{codedefinition}
An $(M,K,n,\epsilon,\sigma,\delta)$ code for
\eqref{channelmodel0} and \eqref{channelmodel1}
consists of an encoder and a decoder
\begin{align*}
 f&:\langle K\rangle\times\langle M\rangle
 \to\mathbb R^n,\\
 \phi&:\mathbb R^n\times\langle K\rangle
 \to\langle M\rangle.
\end{align*}
With $\bm x_{sw}=f(s,w)$ and
$\hat W=\phi(\bm Y,S)$, the code satisfies
\begin{equation}\label{errordefintion}
 \max_{\substack{s\in\langle K\rangle\\
                 w\in\langle M\rangle}}
 \mathbb P(\hat W\neq w\mid S=s,W=w)
 \leq\epsilon
\end{equation}
and
\begin{equation}\label{covertconstraint}
 \mathbb D\!\left(
 Q_{\bm Z,\bm c}^{(n)}
 \middle\|Q_{\bm Z,0}^{(n)}
 \right)\leq\delta,
\end{equation}
where the induced output density is
\begin{equation}\label{mixtureGaussian}
 Q_{\bm Z,\bm c}^{(n)}(\bm z)
 =
 \frac{1}{MK}\sum_{s=1}^{K}\sum_{w=1}^{M}
 W_2^n(\bm z\mid\bm x_{sw}),
\end{equation}
and $Q_{\bm Z,0}^{(n)}
=\mathcal N(\bm0,\sigma^2\bm I_n)$ is Willie's
output distribution when Alice is silent.
\end{Definition}

\begin{Definition}\label{Awgncovertcodes}
An $(M,K,n,\epsilon,\Psi,\sigma,\delta)_m$ code
satisfies Definition~\ref{codedefinition} and
the maximal-power constraint
\begin{equation}\label{maximalpower1}
 \|\bm x_{sw}\|^2\leq n\Psi,
 \qquad \forall\,s,w.
\end{equation}
An $(M,K,n,\epsilon,\Psi,\sigma,\delta)_a$ code
instead satisfies the same definition with the
average-power constraint for each key:
\begin{equation}\label{averagepower1}
 \frac{1}{M}\sum_{w=1}^{M}\|\bm x_{sw}\|^2
 \leq n\Psi,
 \qquad \forall\,s.
\end{equation}
\end{Definition}

Let $M_m^*(n,\epsilon,\Psi)$ and
$M_a^*(n,\epsilon,\Psi)$ denote the supremal message
sizes of conventional AWGN codes under maximal-
and average-power constraints, respectively, with
maximal error probability at most $\epsilon$.
Their covert counterparts are
$M_m^*(n,\epsilon,\Psi,\sigma,\delta)$ and
$M_a^*(n,\epsilon,\Psi,\sigma,\delta)$.

Without an additional power constraint, define
\begin{equation}
 M^*(n,\epsilon,\sigma,\delta)
 :=\sup\{M\in\mathbb N:
 M\text{ is achievable}\},
\end{equation}
where achievability requires an
$(M,K,n,\epsilon,\sigma,\delta)$ code for some
finite $K$. The power-constrained covert quantities
additionally require \eqref{maximalpower1} or
\eqref{averagepower1}. In all three cases, $K$ is
unrestricted and may depend on $n$. Note that our reliability and covertness metrics are maximal
decoding error probability \cite{Yu0,Yu1,M.Tahmasbi2}
and KL divergence \cite{Boulat A,Ligong Wang},
respectively.

The following relations hold.
\begin{Lemma}\label{powercodes}
\begin{align}
&M_m^*(n,\epsilon,\Psi,\sigma,\delta)
\leq M_m^*(n,\epsilon,\Psi)
\leq M_a^*(n,\epsilon,\Psi),
\label{relation0}\\
&M_m^*(n,\epsilon,\Psi,\sigma,\delta)
\leq M^*(n,\epsilon,\sigma,\delta),
\label{relation1}\\
&M_m^*(n,\epsilon,\Psi,\sigma,\delta)
\leq M_a^*(n,\epsilon,\Psi,\sigma,\delta)
\leq M^*(n,\epsilon,\sigma,\delta).
\label{relation2}
\end{align}
\end{Lemma}

\noindent\emph{Proof:}
See the Appendix.

Our objective is to derive achievable lower
bounds and converse upper bounds on
$M^*(n,\epsilon,\sigma,\delta)$.
The achievability analysis also provides
a sufficient secret-key size $K_n$.

\subsection{Main Results and Outline of The Proof}
In this section,  we provide the main results and a road map on the proof of the achievability and converse bounds.

\begin{Theorem}\label{achievability0}
For the AWGN covert channels in
\eqref{channelmodel0} and \eqref{channelmodel1},
fix $\mu\in(1/2,1)$, $\delta>0$, $\sigma>0$, $\epsilon\in(0,1/2)$, and define
\begin{align}
A_\mu
&\triangleq \mu\sigma^2\sqrt{\delta\log e},
\\
B_\mu
&\triangleq \sigma\sqrt{\frac{2}{\mu}}\,
\delta^{1/4}(\log e)^{3/4}.
\end{align}
Then, for sufficiently large $n$, the achievable message size satisfies
\begin{equation}\label{eq:overview-achievability}
\begin{aligned}
\log M^*(n,\epsilon,\sigma,\delta)
\geq {}&
A_\mu\sqrt n\\
&+B_\mu n^{1/4}\Phi^{-1}(\epsilon)
+o(n^{1/4}).
\end{aligned}
\end{equation}
Moreover, there exists a sequence of codes attaining the stated achievable throughput with a secret-key size $K_n$ satisfying
\begin{equation}
\log K_n=
\begin{cases}
\bar{A}_{\mu}\sqrt{n}+Bn^{1/4}+o(n^{1/4}),
& \bar{A}_{\mu}>0,\\[3pt]
Bn^{1/4}+o(n^{1/4}),
& \bar{A}_{\mu}=0,\\[3pt]
0 \quad
& \bar{A}_{\mu}<0,
\end{cases}
\end{equation}
where
\begin{align}
\bar{A}_{\mu}
&\triangleq
\frac{1-\mu^2\sigma^2}{\mu}\sqrt{\delta\log e},\\
B
&\triangleq
\sigma\sqrt{\frac{2}{\mu}}\,
\delta^{1/4}(\log e)^{3/4}\Phi^{-1}(1-\epsilon).
\end{align}
The three cases correspond to $\sigma^2<\mu^{-2}$, $\sigma^2=\mu^{-2}$, and $\sigma^2>\mu^{-2}$.
Furthermore, the converse bound is
\begin{equation}\label{eq:overview-converse}
\begin{aligned}
\log M^*(n,\epsilon,\sigma,\delta)
\leq {}&
\frac{A_\mu}{\mu}\sqrt n\\
&+\sqrt{\mu}B_\mu n^{1/4}
\Phi^{-1}(\epsilon)
+O(\log n).
\end{aligned}
\end{equation}
\end{Theorem}
\begin{Remark}
The achievability and converse bounds in
Theorem \ref{achievability0} have the same
first- and second-order scalings,
$O(\sqrt{n})$ and $O(n^{1/4})$,
but different coefficients for fixed
$\mu\in(1/2,1)$.
In particular, their first-order coefficients
are $\mu\sigma^2\sqrt{\delta\log e}$
and $\sigma^2\sqrt{\delta\log e}$,
respectively.
Thus, the theorem provides explicit
achievable asymptotic terms and a converse
bound, rather than an exact characterization.
Although the gap decreases as $\mu\to1$,
allowing $\mu=\mu_n\to1$ requires uniform
reliability, resolvability, and expurgation
estimates, which remain beyond the scope
of this work.
\end{Remark}

The proof of Theorem \ref{achievability0}
proceeds as follows.

\emph{Achievability:}
Section \ref{oneshotachi} introduces truncated
Gaussian random coding and derives the
decoding error bounds.
Section \ref{Resolvability} establishes an
ensemble-average KL-divergence bound.
Section \ref{Mainachievability} then applies
concentration and expurgation arguments
to establish the existence of codebooks
satisfying both maximal reliability and
covertness constraints.

\emph{Converse:}
Lemma \ref{conversecorollary} derives a
power bound from the covertness constraint.
Combined with the pigeonhole principle,
Corollary \ref{conversecorollary2} yields
the converse bound.

Theorem~1 establishes explicit achievable first-
and second-order throughput terms, together with
a converse bound, for covert communication over
AWGN channels. Both bounds exhibit a leading
term of order $O(\sqrt{n})$ and a second-order
term of order $O(n^{1/4})$, with coefficients
depending on the decoding error probability
$\epsilon$, adversary noise variance $\sigma^2$,
and covertness constraint $\delta$.
The achievable coefficients additionally depend
on the truncation parameter $\mu$, reflecting
the effect of the truncated Gaussian input
distribution. Since the leading achievable
and converse coefficients do not coincide
for fixed $\mu<1$, these results provide
quantitative achievability and converse bounds
rather than an exact asymptotic characterization
of maximal covert throughput.
The effects of $\mu$ and $\sigma^2$ on achievable
throughput and sufficient secret-key size are
examined in the following sections.

\subsection{Related Distributions}\label{preliminarydistributions}

We introduce the distributions used in the
reliability and resolvability analyses.

\emph{1) Transmitter:}
Let $P_n^{(n)}$ be the Gaussian distribution
with density
\begin{equation}\label{gdensity0}
g^{(n)}(\bm{x})
=
\frac{\exp\left(
-\frac{\|\bm{x}\|^2}{2\mu\Psi(n)}
\right)}
{(2\pi\mu\Psi(n))^{n/2}}.
\end{equation}
The truncated distribution $\bar{P}_n^{(n)}$
has density
\begin{equation}\label{fdensity}
\bar{g}^{(n)}(\bm{x})
=
\frac{g^{(n)}(\bm{x})}{\Delta}
\mathbf{1}\{\bm{x}\in\mathcal{A}_n\},
\end{equation}
where
\begin{align}
\mathcal{A}_n
&=\left\{\bm{x}:
\mu\sqrt{n\Psi(n)}
\leq\|\bm{x}\|
\leq\sqrt{n\Psi(n)}\right\},\\
\Delta&=P_n^{(n)}(\mathcal{A}_n),\label{Deltadefintion}
\qquad 0<\mu<1.
\end{align}
The function $\Psi(n)$ specifies the blocklength-dependent
power scale, while $\mu\in(1/2,1)$ is a fixed
design parameter determining the Gaussian reference
variance $\mu\Psi(n)$ and the truncation shell $\mathcal{A}_n$.

\emph{2) Legitimate receiver (Bob):}
The noise distribution is
$Q_{\bm{Y},0}^{(n)}
=\mathcal{N}(\bm{0},\bm{I}_n)$,
with density $f_0^{(n)}$.
The outputs induced by $P_n^{(n)}$
and $\bar{P}_n^{(n)}$ are
$Q_{\bm{Y},1}^{(n)}$ and
$\bar{Q}_{\bm{Y},1}^{(n)}$, with densities
\begin{equation}
f_1^{(n)}=g^{(n)}*f_0^{(n)},
\quad
\bar{f}_1^{(n)}=\bar{g}^{(n)}*f_0^{(n)}.
\end{equation}

\emph{3) Adversary (Willie):}
The noise distribution is
$Q_{\bm{Z},0}^{(n)}
=\mathcal{N}(\bm{0},\sigma^2\bm{I}_n)$,
with density $h_0^{(n)}$.
The outputs induced by $P_n^{(n)}$
and $\bar{P}_n^{(n)}$ have densities
\begin{equation}
h_1^{(n)}=g^{(n)}*h_0^{(n)},
\quad
\bar{h}_1^{(n)}=\bar{g}^{(n)}*h_0^{(n)}.
\end{equation}
The codebook-induced distribution
$Q_{\bm{Z},\bm{c}}^{(n)}$
is defined in (\ref{mixtureGaussian}).

The following results quantify the closeness
between $P_n^{(n)}$ and $\bar{P}_n^{(n)}$.

\begin{figure}
\includegraphics[width=3in]{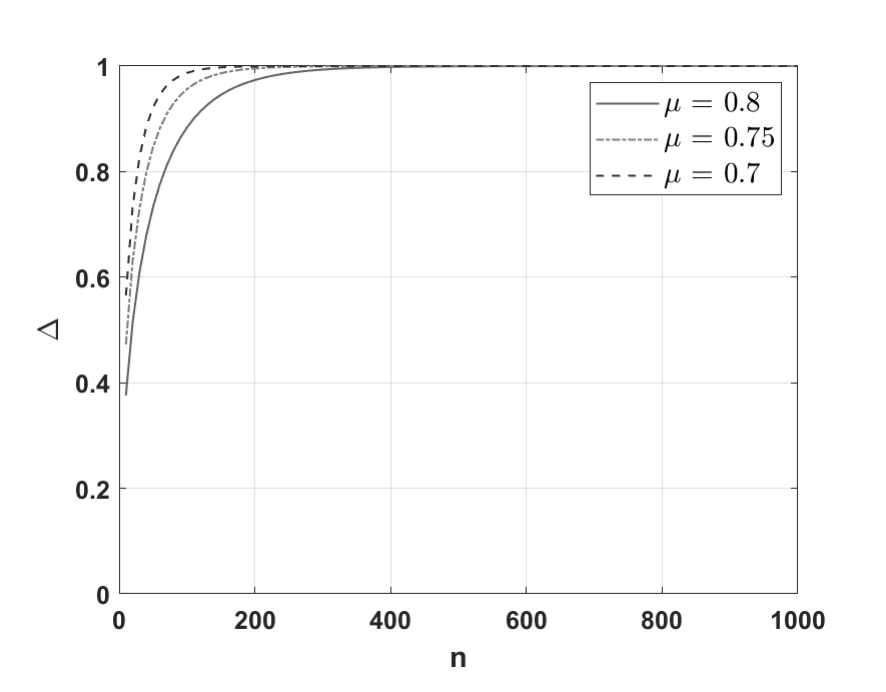}
\caption{$\Delta$ tends to $1$ with proper chosen $\mu$, which is the numerical results from (\ref{constantmu}).}\label{Fig2}
\end{figure}

\begin{Lemma}\label{inputdistributionapprox101}
The KL divergence and TVD between $\bar{P}_n^{(n)}$ and $P_n^{(n)}$ are
\begin{equation}\label{KLclose1}
\mathbb{D}(\bar{P}_n^{(n)}, P_n^{(n)}) = - \log \Delta,
\end{equation}
\begin{equation}\label{TVDclose1}
\mathbb{V}(\bar{P}_n^{(n)}, P_n^{(n)}) = 1- \Delta.
\end{equation}
with 
\begin{equation}\label{constantmu}
\Delta =  \frac{\gamma(n/2, n/2\mu)- \gamma(n/2, n\mu/2)}{\Gamma(n/2)},
\end{equation}
where $\gamma(a,z)$ is incomplete gamma function defined as follows
 \begin{equation}\label{Delta}
 \gamma(a, z)= \int_0^{z}e^{-t}t^{a-1}dt.
 \end{equation}
\end{Lemma}
The proof can be found in the Appendix. With $\frac{1}{2} < \mu < 1$ being a constant, $\Delta \rightarrow 1$ as $n\rightarrow \infty$ exponentially and the rapidity depends only on $\mu$. Please see Fig.\ref{Fig2} for reference, and the details of the analysis has been presented in Section IV-A of \cite{Yu1}.


 For the terms $\frac{\bar{Q}^{(n)}_{\bm{Y},1}(\bm{y})} {Q^{(n)}_{\bm{Y},1}(\bm{y})}$ and  $\frac{\bar{Q}^{(n)}_{\bm{Z},1}(\bm{z})}{Q^{(n)}_{\bm{Z},1}(\bm{z})}$, 
we have the following bounds for all $n > 0$.
\begin{Lemma}\label{lemmanecessary1}
For every $\bm y,\bm z\in\mathbb R^n$,
\begin{align}\label{densityratiobounds}
0<\frac{\bar f_1^{(n)}(\bm y)}{f_1^{(n)}(\bm y)}\leq\frac1\Delta,
\qquad
0<\frac{\bar h_1^{(n)}(\bm z)}{h_1^{(n)}(\bm z)}\leq\frac1\Delta.
\end{align}
Moreover,
\begin{align}\label{inequalities1}
\mathbb{V}(\bar Q_{Y,1}^{(n)},Q_{Y,1}^{(n)})\leq1-\Delta, \, \ \,
\mathbb{D}(\bar Q_{Y,1}^{(n)}\|Q_{Y,1}^{(n)})\leq-\log\Delta.
\end{align}
\end{Lemma}
\noindent\emph{Proof:} See the Appendix.


\begin{Remark}
For simplicity, we use distribution notation
also for the corresponding densities when
their arguments are specified.
\end{Remark}

\section{Throughput Analysis of Covert Communications Over AWGN channels}\label{achievabilitysection} 
\subsection{Random Coding, Decoding and Reliability Analysis}\label{oneshotachi}

This section establishes one-shot achievability
bounds using truncated Gaussian random coding
and threshold decoding. Following
\cite{M.Tahmasbi2}, we separately analyze
the Type-I and Type-II decoding errors.
A suitable decoding threshold ensures a
uniform bound on $\varepsilon_{sw}^{(1)}$
over the truncated codeword region, whereas
the ensemble-average bound on
$\varepsilon_{sw}^{(2)}$ is subsequently
strengthened through expurgation.

Before a realization is selected, the codebook family
$\mathcal C$ is random. Alice generates $K$ subcodes,
each containing $M$ codewords, by independently
sampling all $MK$ codewords
$\{\bm{x}_{sw}:s\in\langle K\rangle,
w\in\langle M\rangle\}$ from $\bar P_n^{(n)}$.
We denote a realization of $\mathcal C$ by $\bm c$.
The realized codebook family is publicly known,
while the pre-shared secret key $S=s$ selects
the $s$-th subcode used for transmission.
Given the received vector $\bm{y}$,
Bob declares $\widehat{W}=w$ if $w$ is
the unique index satisfying
\begin{equation}\label{decoder}
\log
\frac{W_1^n(\bm{y}|\bm{x}_{sw})}
{Q_{\bm{Y},1}^{(n)}(\bm{y})}
>\gamma.
\end{equation}
Otherwise, an error is declared.
This decoder follows the threshold-decoding
framework of \cite{Polyanskiy,M.Tahmasbi2}.


The conditional decoding error probability
is bounded by the sum of two terms:
\begin{align}
\varepsilon_{sw}^{(1)}
&=
\mathbb{P}\left(
\log\frac{W_1^n(\bm{Y}|\bm{x}_{sw})}
{Q_{\bm{Y},1}^{(n)}(\bm{Y})}
\leq\gamma
\right),
\label{conditionerror1}\\
\varepsilon_{sw}^{(2)}
&=
\mathbb{P}\left(
\exists w'\neq w:
\log\frac{W_1^n(\bm{Y}|\bm{x}_{sw'})}
{Q_{\bm{Y},1}^{(n)}(\bm{Y})}
>\gamma
\right),
\label{conditionerror2}
\end{align}
where the probabilities are evaluated
conditioned on the transmission of
$\bm{x}_{sw}$.
Here, $Q_{\bm{Y},1}^{(n)}$ is induced by
the Gaussian distribution $P_n^{(n)}$,
rather than the truncated distribution
$\bar{P}_n^{(n)}$.
The codebook $\mathcal C$ is randomly generated.
The probabilities $\varepsilon_{sw}^{(1)}$ and
$\varepsilon_{sw}^{(2)}$ are evaluated conditional
on its realization and the transmission of
$\bm{x}_{sw}$. For each realization, the probabilities
are taken over Bob's channel noise. Consequently,
both error probabilities are random variables
with respect to the codebook ensemble.

The following lemma establishes a uniform
bound on $\varepsilon_{sw}^{(1)}$ using
the Berry--Esseen theorem \cite{Feller}.
Unlike the equal-power settings in
\cite{Polyanskiy,Tan0}, the codeword energies
vary over
$[\mu^2n\Psi(n),n\Psi(n)]$.
We exploit spherical symmetry at each
fixed radius and select a common decoding
threshold to bound the Type-I error
uniformly over the truncated region.

\begin{Lemma}\label{decodingerrortype1}
Let
\[
 \beta_n:=\frac1{\sqrt n}\sup_{\Upsilon\in[\mu^2\Psi(n),\Psi(n)]}
 B(\Upsilon,\mu\Psi(n)).
\]
For any $\epsilon_1$ satisfying $0<\epsilon_1-\beta_n<1/2$, choose the common decoding threshold $\gamma$ as follows.
\begin{align}\label{gammadefintion}
\gamma  =  & n\mathsf{C}(\mu, \Psi(n))  +  n\bar{\mathsf{C}}(\mu^2 \Psi(n),\mu\Psi(n)) \nonumber \\
 + & \sqrt{n\mathsf{V}(\Psi(n),\mu\Psi(n)) } \nonumber \\
 \times & \Phi^{-1}\bigg(\epsilon_1 - \frac{\underset{\Upsilon \in [\mu^2 \Psi(n), \Psi(n)]}{\max} B(\Upsilon, \mu\Psi(n))}{\sqrt{n}}\bigg),
 \end{align}
where the quantities in (\ref{gammadefintion}) are defined as follows. 
 \begin{equation}\label{capacitymain1}
\mathsf{C}(\mu,\Psi(n)) := \frac{1}{2}\log (1 + \mu \Psi(n)),
\end{equation} 
\begin{equation}\label{capacitymain2}
\bar{\mathsf{C}}(\mu^2\Psi(n),\mu\Psi(n)) := \frac{\mu^2 \Psi(n) - \mu \Psi(n)}{2(1+\mu \Psi(n))} \log e.
\end{equation}
In addition, for $\Upsilon \in [\mu^2 \Psi(n), \Psi(n)]$
\begin{align}
S_i(\Upsilon,\mu\Psi(n))
&:=\frac{\log e}{2(1+\mu\Psi(n))}
\big[\Upsilon-\mu\Psi(n)\xi_i^2+2\sqrt\Upsilon\xi_i\big], \nonumber\\
\bar{\mathsf C}(\Upsilon,\mu\Psi(n))
&:=\frac{\Upsilon-\mu\Psi(n)}{2(1+\mu\Psi(n))}\log e,
\qquad \xi_i\stackrel{\rm iid}{\sim}\mathcal N(0,1),\nonumber
\end{align}
\begin{align}
\mathsf{T}&\left[S_i(\Upsilon,\mu\Psi(n))\right] \nonumber\\
:= & \frac{1}{n} \sum_{i=1}^n\mathbb{E}\left[|S_i(\Upsilon,\mu\Psi(n)) - \bar{\mathsf{C}}(\Upsilon,\mu\Psi(n)) |^3\right],
\end{align}
\begin{equation}\label{eq:conditional-dispersion}
\mathsf V(\Upsilon,\mu\Psi(n))
:=
\frac{2\Upsilon+\mu^2\Psi(n)^2}
     {2(1+\mu\Psi(n))^2}(\log e)^2, 
\end{equation}
\begin{equation}\label{Bexpression}
B(\Upsilon, \mu\Psi(n)) := \frac{6\mathsf{T}\left[S_i(\Upsilon,\mu\Psi(n))\right]}{\mathsf{V}\big(\Upsilon,\mu\Psi(n)\big)^{\frac{3}{2}}}.
\end{equation}
 and $\Phi$ is the cumulative distribution function of standard normal random variable. 
Then we have
 \begin{equation}\label{maximalerrortype1}
\underset{s,w}{\max}\, \ \, \varepsilon_{sw}^{(1)} \leq \epsilon_1.
\end{equation} 
\end{Lemma}
\noindent\emph{Proof:} See the Appendix.
\begin{figure}
\includegraphics[width=3in]{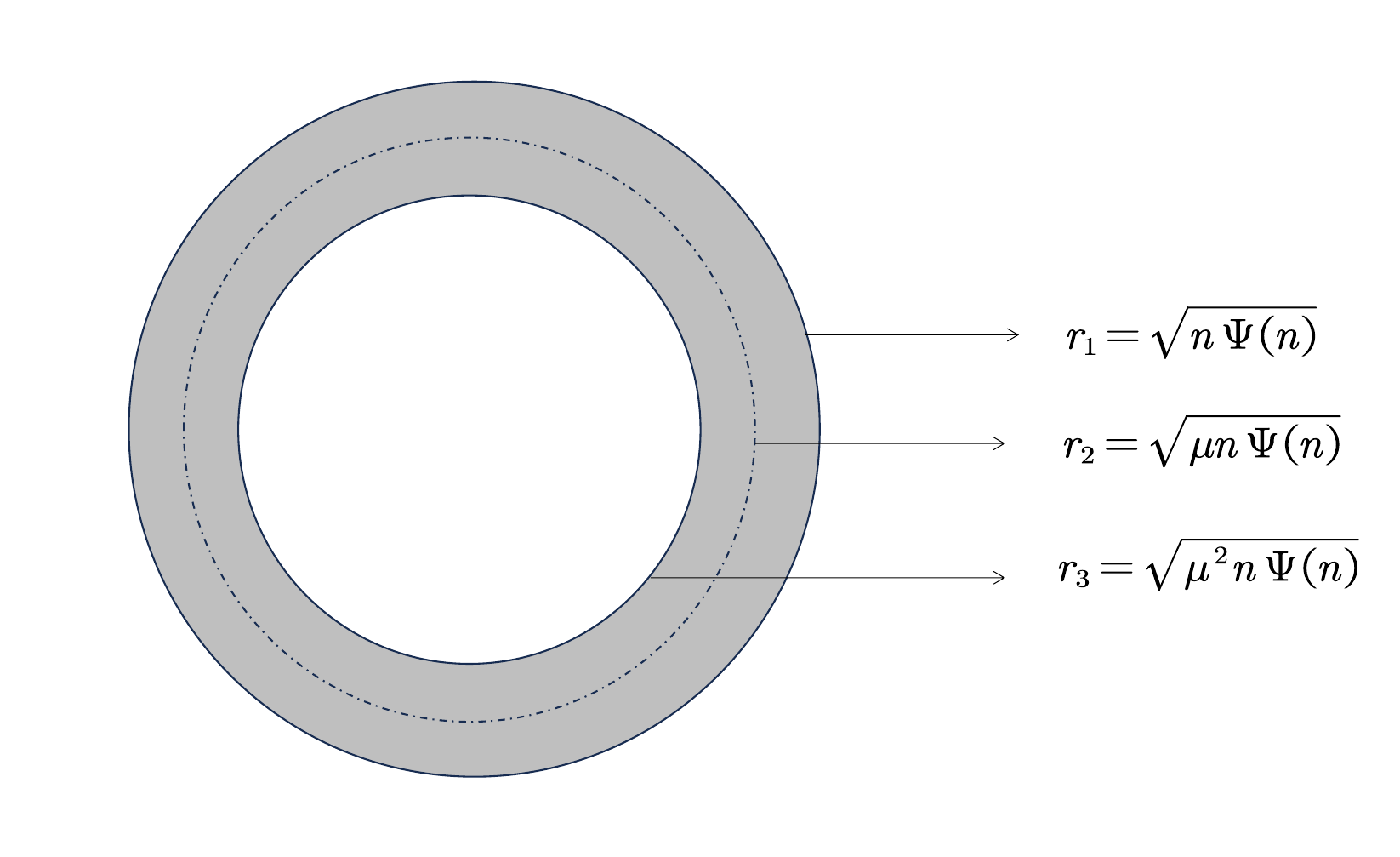}
\caption{The codewords lie in the gray region.}\label{Fig222}
\end{figure}

\begin{Remark}\label{firsttypeerrorremark1}
Unlike the constant-energy codewords considered in
\cite{Polyanskiy,Tan0}, our codewords are generated
from a truncated Gaussian distribution and lie in
the spherical shell
$\mathcal{A}_n$, as illustrated in
Fig.~\ref{Fig222}.
To bound the Type-I decoding error probability,
we condition on the codeword energy
$\|\bm{x}\|^2=n\Upsilon$ and exploit spherical
symmetry. The lower truncation radius facilitates
the selection of a common decoding threshold
$\gamma$ for all admissible codewords.
Consequently, the normal approximation involves
the additional term
$\bar{\mathsf{C}}(\Upsilon,\mu\Psi(n))$,
reflecting the variation in codeword energy.
A smaller $\mu$ enlarges this variation and may
lead to a more restrictive codebook-size condition
to satisfy the reliability constraint.
\end{Remark}

\begin{Remark}
The RCU analysis in \cite{Tan0} relies on a uniform
bound on the ratio between the induced and auxiliary
output densities over a high-probability set
$\mathcal{F}$.
For our truncated Gaussian construction, a
corresponding set is
\begin{align}
\mathcal{F}=\bigg\{\bm{y}\in\mathbb{R}^n:
&\,1+\mu^2\Psi(n)-\tau_n
\nonumber\\
&\leq\frac{\|\bm{y}\|^2}{n}
\leq1+\Psi(n)+\tau_n
\bigg\}.
\end{align}
However, the uniform density-ratio bound in
\cite{Tan0} does not directly extend to this
setting because of the varying codeword energies.
We therefore bound the Type-I decoding error
directly by conditioning on the codeword energy.
\end{Remark}

The preceding lemma establishes a uniform Type-I error bound, $\max_{s,w}\varepsilon_{sw}^{(1)}\leq\epsilon_1$. For the Type-II error, we first derive an average bound in Lemma \ref{averagesecondtypeerror} and then apply expurgation to remove codewords with large error probabilities. The following lemma extends \cite[Lemma 1]{M.Tahmasbi2} to the AWGN channel by replacing summations with integrals and adopting base-2 logarithms.
\begin{Lemma}\label{Blochlemma}
For the AWGN channel (\ref{channelmodel0}), let $P_{X^n}$, $P_{Y^n}$ and $Q_{Y^n}$ be distributions on $\mathbb{R}^n$, where $P_{Y^n}$ is the induced distribution of $P_{X^n}$. If we choose $\bm{x}_1, \cdots, \bm{x}_M$ independently according to $P_{X^n}$, then by the decoding rule (\ref{decoder}), for all $\gamma \in \mathbb{R}$ and $w\in \langle M\rangle$
\begin{equation}
\mathbb{E}\left(\varepsilon_{sw}^{(2)}\right) \leq  M\cdot2^{-\gamma} \mathbb{E}_{P_{Y^n}}\bigg(\frac{P_{Y^n}}{ Q_{Y^n}} \bigg).
\end{equation}
\end{Lemma}
Let $P_{X^n}$ be the truncation Gaussian distribution $\bar{P}_n^{(n)}$, and its induced distribution will be $\bar{Q}^{(n)}_{\bm{Y},1}$. Using the fixed distribution $Q^{(n)}_{\bm{Y},1}$ for $Q_{Y^n}$.  Lemma \ref{Blochlemma} turns out in the following form in our setting.
\begin{equation}\label{secondkinderror}
\begin{split}
 \mathbb{E}_{\mathcal{C}}\left(\epsilon_{sw}^{(2)}\right) 
\leq & M_n\cdot2^{-\gamma} \mathbb{E}_{\bar{Q}^{(n)}_{\bm{Y},1}}\bigg( \frac{\bar{Q}^{(n)}_{\bm{Y},1}} {Q^{(n)}_{\bm{Y},1}}\bigg) \\
\leq & M_n\cdot2^{-\gamma} \mathbb{E}_{\bar{Q}^{(n)}_{\bm{Y},1}}\bigg( \frac{\bar{f}_1^{(n)}(\bm{y})}{ f_1^{(n)}(\bm{y})}\bigg)
\end{split}
\end{equation}
The above upper bound on the average of the error probability of the second type can be further elaborated if we choose a proper parameter $M$. 

\begin{Lemma}\label{averagesecondtypeerror}
If we choose $M_n=\lfloor 2^{\gamma-c_0\log n}\rfloor$
\begin{align}\label{averageerrortype2}
\mathbb{E}_{\mathcal C}[\varepsilon_{sw}^{(2)}]
 \leq & M_n2^{-\gamma}\mathbb{E}_{\bar Q_{\bm Y,1}^{(n)}} 
\left[\frac{\bar f_1^{(n)}(\bm Y)}{f_1^{(n)}(\bm Y)}\right]\nonumber\\
&\leq n^{-c_0}\mathbb{E}_{\bar Q_{\bm Y,1}^{(n)}}
\left[\frac{\bar f_1^{(n)}(\bm Y)}{f_1^{(n)}(\bm Y)}\right] \nonumber\\
&\leq\frac{1}{\Delta n^{c_0}}.
\end{align}
\end{Lemma}

The proof follows directly from (\ref{secondkinderror})
and (\ref{densityratiobounds}). Since the bound applies
only to the average Type-II error probability,
expurgation will be employed in Theorem
\ref{finiteachievabilty} to ensure that
$\max_{s,w}(\varepsilon_{sw}^{(1)}
+\varepsilon_{sw}^{(2)})$ satisfies the
reliability constraint.

\begin{Remark}
The average Type-II error bound in
(\ref{averageerrortype2}) depends on the codebook
size and the truncation effect parameter $\Delta$.
\end{Remark}


\subsection{Resolvability Analysis}\label{Resolvability}
This section presents a one-shot resolvability analysis
for covert communication over AWGN channels.
For a realization $\bm c$ of the random codebook
family $\mathcal C$, covertness is measured by
$\mathbb D(Q_{\bm Z,\bm c}^{(n)}
\|Q_{\bm Z,0}^{(n)})$.
Before fixing the codebook, this divergence is
a random variable through $\mathcal C$ and is denoted as $\mathbb{D}\bigl(
Q^{(n)}_{\bm{Z},\mathcal{C}}
\|Q^{(n)}_{\bm{Z},0}
\bigr)$. 
Unlike the discrete-channel setting in
\cite{M.Tahmasbi2}, the modified triangle
inequality used therein does not directly
extend to our setting. Instead, we use
the following decomposition:
\begin{align}\label{decomposition}
&\mathbb{D}\bigl(
Q^{(n)}_{\bm{Z},\mathcal{C}}
\|Q^{(n)}_{\bm{Z},0}
\bigr)\\
&=
\mathbb{D}\bigl(
Q^{(n)}_{\bm{Z},\mathcal{C}}
\|Q^{(n)}_{\bm{Z},1}
\bigr)
+
\int_{\mathbb{R}^n}
Q^{(n)}_{\bm{Z},\mathcal{C}}(\bm{z})
\log
\frac{Q^{(n)}_{\bm{Z},1}(\bm{z})}
{Q^{(n)}_{\bm{Z},0}(\bm{z})}
\,d\bm{z}.\nonumber
\end{align}
Taking the expectation over the random codebook
$\mathcal{C}$ yields
\begin{align}\label{expection1}
&\mathbb{E}_{\mathcal{C}}
\left[
\mathbb{D}\bigl(
Q^{(n)}_{\bm{Z},\mathcal{C}}
\|Q^{(n)}_{\bm{Z},0}
\bigr)
\right]
\nonumber\\
&=
\mathbb{E}_{\mathcal{C}}
\left[
\mathbb{D}\bigl(
Q^{(n)}_{\bm{Z},\mathcal{C}}
\|Q^{(n)}_{\bm{Z},1}
\bigr)
\right]
\nonumber\\
&\quad+
\mathbb{E}_{\mathcal{C}}
\left[
\int_{\mathbb{R}^n}
Q^{(n)}_{\bm{Z},\mathcal{C}}(\bm{z})
\log
\frac{Q^{(n)}_{\bm{Z},1}(\bm{z})}
{Q^{(n)}_{\bm{Z},0}(\bm{z})}
\,d\bm{z}
\right].
\end{align}
We bound the two terms in (\ref{expection1})
separately. The following lemma bounds the
first term, with further analysis provided
in Corollary \ref{corollary3}.

\begin{Lemma}\label{initalbound1}
\begin{align}\label{expectation2terms}
\mathbb{E}_{\mathcal{C}}\left[\mathbb{D}(Q^{(n)}_{\bm{Z},\mathcal{C}}\| Q^{(n)}_{\bm{Z},1})\right] 
\leq &-\log\Delta   \\ +
\mathbb E_{\bar P_n^{(n)}W_2^n}
&\log\left(1+\frac{W_2^n(\bm Z\mid\bm X)}{MK\,Q_{\bm{Z},1}^{(n)}(\bm Z)}\right).\nonumber
\end{align}
\end{Lemma}
\noindent\emph{Proof:} See the Appendix.

Now we consider the second term in (\ref{expectation2terms}). First, let's define the event
\[
\mathcal{E}=\left\{(\bm x,\bm z):
\log\frac{W_2^n(\bm z\mid\bm x)}{Q_{Z,1}^{(n)}(\bm z)}\leq\gamma_0\right\}.
\]
where $\gamma_0$ is a function of $\mu, \Psi(n)$ and $\sigma^2$. 
Then, some technical preparations are provided in Lemma \ref{Lemmalabel1} and Lemma \ref{eventcap}. 

\begin{Lemma}\label{Lemmalabel1}
For every $\bm x\in\mathcal A_n$ and $\bm z\in\mathbb R^n$, the following inequality holds
\begin{equation}
\frac{W_2^n(\bm{z}|\bm{x})}{{Q}^{(n)}_{\bm{Z},1}(\bm{z})} \leq \bigg(\frac{\sigma^2 + \mu \Psi(n)}{\sigma^2}\bigg)^{\frac{n}{2}} e^{ \frac{n}{2\mu }}.
\end{equation}
\end{Lemma}
\noindent\emph{Proof:} See the Appendix.

Next, we divide the product space $\bm{X} \times \bm{Z}$ into two parts $\mathcal{E}$ and $\mathcal{E}^c$. For the event $\mathcal{E}^c  = \{(\bm{x},\bm{z}): \log \frac{W_2^n(\bm{z}|\bm{x})}{Q^{(n)}_{\bm{Z},1}(\bm{z})} > \gamma_0\}$, the following Lemma  \ref{eventcap} provides an upper bound on its probability with proper chosen $\gamma_0$. Similar with the proof of Lemma \ref{decodingerrortype1}, our strategy here is separately considering the codewords with equal power and choosing an universal threshold $\gamma_0$. As a consequence, the sphere symmetry can be leveraged to show that the probability of the event $\{\bm{z}:\frac{W_2^n(\bm{z}|\bm{x})}{Q^{(n)}_{\bm{Z},1}(\bm{z})} > \gamma_0\}$ for each $\bm{x}$ has large-deviation bounds. 
\begin{Lemma}\label{eventcap}
If we choose \begin{align}\label{gamma0definition}
\gamma_0=n\widetilde{\mathsf{C}}(\mu,\Psi,\sigma^2)-n\widehat{\mathsf{C}}(\Psi(n),\mu\Psi(n),\sigma^2),
\end{align}
where
\begin{equation}\label{secondcapacityformula}
\widetilde{\mathsf{C}}(\mu,\Psi,\sigma^2):= \frac{1}{2}\log (1 + \frac{\mu \Psi(n)}{\sigma^2})
\end{equation} 
\begin{equation}
\widehat{\mathsf{C}}(\Psi(n),\mu\Psi(n),\sigma^2):= \frac{\log e}{2(\sigma^2 + \mu \Psi(n))} ( \mu\Psi(n) - \Psi(n)),
\end{equation}
In addition, let $h=\frac{1+\mu}{2\mu}>1$ and 
$c_E=\frac{h-1-\ln h}{2}$.
 Then
\begin{align}\label{eq:zero-tail}
 p_n^{(0)}:= \mathbb{P}(\mathcal{E}^c)
 &\leq\Delta^{-1}e^{-c_En}+e^{-an}+e^{-bn\Psi(n)}.
\end{align}
with $a=\frac{(1-\mu)^2}{64\mu^2},\, \,
 b=\frac{(1-\mu)^2}{128\sigma^2}$.
\end{Lemma}
\noindent\emph{Proof:} See the Appendix.

Lemma \ref{eventcap} bounds the information-density tail under the joint truncated-input and channel law. Its proof controls the conditional tail uniformly on a typical input-energy region and bounds the probability of the remaining input energies separately. 
\begin{Lemma}\label{firsttermupperbound}
\begin{align}
&\mathbb{E}_{
W_2^n(\bm{Z}|\bm{X})
\bar{P}_n^{(n)}(\bm{X})
}
\left[
\log\left(
1+\frac{
W_2^n(\bm{Z}|\bm{X})
}{
MKQ_{\bm{Z},1}^{(n)}(\bm{Z})
}
\right)
\right]
\nonumber\\
&\leq
\log\left(1+\frac{2^{\gamma_0}}{MK}\right)+dn p_n^{(0)},
\end{align}
where $\gamma_0$ is defined in
(\ref{gamma0definition}) and $d>0$
is a constant.
\end{Lemma}

\noindent\emph{Proof:}
See the Appendix.

From Lemma \ref{initalbound1} and Lemma \ref{firsttermupperbound}, the following conclusion is obvious. 
\begin{Corollary}\label{corollary3}
\begin{align}
\mathbb{E}_{\mathcal{C}}& \left[\mathbb{D}(Q^{(n)}_{\bm{Z},\mathcal{C}}\| Q^{(n)}_{\bm{Z},1})\right] \nonumber\\
\leq & \log\left(1+\frac{2^{\gamma_0}}{MK}\right)+dn p_n^{(0)} -\log \Delta.
\end{align}
\end{Corollary}
At this end, we derive an upper bound on $\mathbb{E}_{\mathcal{C}} \left[\mathbb{D}(Q^{(n)}_{\bm{Z},\mathcal{C}}\| Q^{(n)}_{\bm{Z},0})\right]$ as follows. 
\begin{Theorem}\label{KLexpectation}
For a random code family $\mathcal{C}$ with $MK$ codewords i.i.d sampled from the truncated Gaussian distribution $\bar{P}_n^{(n)}$,
\begin{align}\label{KLthreeterms}
\mathbb{E}_{\mathcal{C}}& \left[\mathbb{D}(Q^{(n)}_{\bm{Z},\mathcal{C}}\| Q^{(n)}_{\bm{Z},0})\right] \\
\leq & \log\left(1+\frac{2^{\gamma_0}}{MK}\right)+dn p_n^{(0)}  \nonumber \\
+ & \mathbb{D}(Q_{Z,1}^{(n)}\|Q_{Z,0}^{(n)})
 + r_n - \log \Delta, \nonumber
\end{align}
where $\gamma_0$ is defined in (\ref{gamma0definition}), $d$ is a positive constant and 
\begin{align}\label{r_ndefintion}
 r_n=\frac{n\mu^2\Psi(n)^2}{2\sigma^2[\sigma^2+\mu\Psi(n)]}
 \frac{1-\Delta}{\Delta}\log e.
 \end{align}
\end{Theorem}
\noindent\emph{Proof:} See the Appendix.

\begin{Remark}
Unlike the sequential random coding scheme in
\cite{Yu1}, our codewords are independently
generated from the truncated Gaussian distribution
$\bar{P}_n^{(n)}$, facilitating resolvability
analysis. Different from \cite{M.Tahmasbi2},
we directly decompose the KL divergence using
the intermediate Gaussian distribution
$Q_{\bm{Z},1}^{(n)}$, rather than relying on
a modified triangle inequality.
\end{Remark}

The bound in (\ref{KLthreeterms}) admits the
following interpretation:
\begin{align}
&\mathbb{E}_{\mathcal{C}}
\left[
\mathbb{D}(Q^{(n)}_{\bm{Z},\mathcal{C}}
\|Q^{(n)}_{\bm{Z},0})
\right]
\nonumber\\
&\leq
\underbrace{
\log\left(1+\frac{2^{\gamma_0}}{MK}\right)
}_{\text{Resolvability}}
+
\underbrace{
dn p_n^{(0)}
}_{\text{Tail penalty}}
\nonumber\\
&\quad+
\underbrace{
\mathbb{D}(Q_{\bm{Z},1}^{(n)}
\|Q_{\bm{Z},0}^{(n)})
}_{\text{Nominal covertness}}
+
\underbrace{
r_n-\log\Delta
}_{\text{Truncation}}.
\label{eq:covertness_bound}
\end{align}

The first two terms characterize the
finite-codebook resolvability penalty,
the third represents the nominal covertness
cost determined by $\Psi(n)$ and $\sigma^2$,
and the last accounts for Gaussian truncation.
\subsection{Existence of the Codebooks and Achievability Bound} \label{Mainachievability}

The preceding sections establish the maximal
Type-I decoding error bound
(\ref{maximalerrortype1}), the average Type-II
error bound (\ref{averageerrortype2}), and the
ensemble-average resolvability bound
(\ref{KLthreeterms}).
We now combine these results with concentration
and expurgation arguments to establish the
existence of a codebook family satisfying both
the maximal reliability and covertness constraints.
Specifically, concentration inequalities control
the deviations from the ensemble-average bounds,
while expurgation removes codewords with large
Type-II decoding error probabilities.
This leads to explicit achievable message and
secret-key sizes.

Our approach follows the general framework of
\cite{M.Tahmasbi2}, but the AWGN setting requires
additional treatment of the adversary's noise
variance $\sigma^2$ and the truncation parameter
$\mu$.
The technical details are provided in the
Appendix.

The following bounded difference inequality is necessary for the application of McDiarmid's Inequality\cite{Wainwright} for the codewords over the AWGN channel, the proof can be found in the Appendix.
\begin{Lemma}\label{KLdifference011}
Let $\mathcal{C}_1 \triangleq \left\{\bm{x}_1, \cdots, \bm{x}_k, \cdots, \bm{x}_M\right\}$ and $\mathcal{C}_2 \triangleq \{\bm{x}_1, \cdots, \bm{x}'_k, \cdots, \bm{x}_M\}$ be two codebooks of codewords with length $n$, which have only one difference on $k$-th codeword. All codewords in $\mathcal{C}_1$ and $\mathcal{C}_2$ lie in the support set of $\bar{P}_n^n$. If $Q^{(n)}_{\bm{Z},\mathcal{C}_1}$ and $Q^{(n)}_{\bm{Z},\mathcal{C}_2}$ are the mixture Gaussian distributions induced by $\mathcal{C}_1$ and $\mathcal{C}_2$ through the channel $W_2$, respectively. The following bound holds for the difference between $\mathbb{D}(Q^{(n)}_{\bm{Z},\mathcal{C}_1}\| Q^{(n)}_{\bm{Z},0})$ and $\mathbb{D}(Q^{(n)}_{\bm{Z},\mathcal{C}_2}\| Q^{(n)}_{\bm{Z},0})$. 
\begin{equation}\label{differencebound0}
\begin{split}
\big| \mathbb{D}(Q^{(n)}_{\bm{Z},\mathcal{C}_1}\| Q^{(n)}_{\bm{Z},0}) & - \mathbb{D}(Q^{(n)}_{\bm{Z},\mathcal{C}_2}\| Q^{(n)}_{\bm{Z},0})\big| 
 \leq   \frac{n}{M} \Theta_n.
\end{split}
\end{equation}
where $\Theta_n = \frac{3\sigma^2+11\Psi(n)}{\sigma^2}\log e
 +2|\log(2\pi\sigma^2)|$.
\end{Lemma} 
Regarding the quantity $\mathbb{D}(Q^{(n)}_{\bm{Z},\mathcal{C}}\|Q^{(n)}_{\bm{Z},0} )$ as a function of the codewords $\bm{x}_{sw},s= 1,\cdots, K,w = 1,\cdots,M$, we apply McDiarmid's Inequality with Lemma \ref{KLdifference011} and obtain the following lemma. As the proof is simple, we omit it here.  
\begin{Lemma}\label{KLdifference1}
Consider the AWGN channel $W_2$ defined in (\ref{channelmodel1}). Let $Q^{(n)}_{\bm{Z},\mathcal{C}}(\bm{z}) = \sum_{s=1}^{K}\sum_{w=1}^{M} \frac{1}{MK} W^n_2(\bm{z}|\bm{x}_{sw})$ be the induced mixture distribution of a random codebook through the channel $W_2$, where the codewords are randomly sampled from $\bar{P}_n^{(n)}$, then for any $\lambda > 0$, the following inequality holds,
\begin{align}
\mathbb{P}\bigg(\mathbb{D}(Q^{(n)}_{\bm{Z},\mathcal{C}}\|Q^{(n)}_{\bm{Z},0} ) - \mathbb{E} \big(\mathbb{D}(Q^{(n)}_{\bm{Z},\mathcal{C}}\|Q^{(n)}_{\bm{Z},0})\big) \geq \lambda \bigg) \nonumber \\
\leq \exp\bigg(- \frac{2MK\lambda^2}{n^2\Theta_n^2}\bigg),
\end{align}
where $\mathbb P$ and $\mathbb E$ are taken over
the random codebook ensemble
$(\bar P_n^{(n)})^{\otimes MK}$, and
\begin{align}\label{Thetaexpression}
\Theta_n = \frac{3\sigma^2+11\Psi(n)}{\sigma^2}\log e
 +2|\log(2\pi\sigma^2)|.
\end{align}
\end{Lemma}
The following lemma, adapted from \cite[Lemma 4]{M.Tahmasbi2}, establishes sufficient conditions for expurgation to preserve covertness and ensure a small average Type-II decoding error for each retained codebook. In contrast with \(M\) and \(K\) as generic
codebook-size parameters, we write \(M_{\rm raw}\) for the number of sampled
codewords and \(M_{\rm fin}\) for the number of retained
codewords corresponding to each key after expurgation. The number of keys remains
\(K\). 
\begin{Lemma}\label{existencesuff}
With AWGN channels $W_1$ and $W_2$ defined in (\ref{channelmodel0}) and (\ref{channelmodel1}), respectively,
choose $M_{\rm raw}$ such that $M_{\rm raw}2^{-\gamma}\leq n^{-c_0}$ with \(\gamma\) from \eqref{gammadefintion} and $c_0 > 0$.  Let
\(\lambda_1>1\), \(\lambda_2\in(0,1)\), and
\(\lambda_3>0\), and define
\begin{align}\label{factM}
M_{\rm fin}:=\lfloor\lambda_2 M_{\rm raw}\rfloor,
\qquad
\widehat\lambda_{2,n}:=\frac{M_{\rm fin}}{M_{\rm raw}}.
\end{align}
The codewords \(\bm x_{sw}\),
\(s\in\langle K\rangle\) and
\(w\in\langle M_{\rm raw}\rangle\),
are sampled independently from
\(\bar P_n^{(n)}\).
For any subset $\mathcal{I} \subset \langle K  \rangle \times\langle M_{\rm raw}\rangle$ with $|\mathcal{I}| = M_{\rm fin}\cdot K $, define a mixture distribution $Q_{\bm{Z},\mathcal{C}}^{\mathcal{I}}$ as
\begin{equation}
Q_{\bm{Z},\mathcal{C}}^{\mathcal{I}} \triangleq \frac{1}{|\mathcal{I}|} \sum_{(s,w) \in \mathcal{I}} W^n_2(\bm{Z}|\bm{x}_{sw}).
\end{equation}
For this mixture distribution and every $\lambda_1 > 1, \lambda_2 \in (0,1), \lambda_3 > 0$, define events
\begin{align}
\mathcal E_1:=
\left\{
\max_{s\in\langle K\rangle}
\frac1{M_{\rm raw}}
\sum_{w=1}^{M_{\rm raw}}\varepsilon_{sw}^{(2)}
\leq\lambda_1n^{-c_0}R_{Y,n}
\right\}.
\end{align} 
with $R_{Y,n} = \mathbb{E}_{\bar Q_{\bm Y,1}^{(n)}}
\left[\frac{\bar f_1^{(n)}(\bm Y)}{f_1^{(n)}(\bm Y)}\right]$;

\begin{align}
\mathcal E_2:=\Big\{&
\text{for every }
\mathcal I\subseteq
\langle K\rangle\times\langle M_{\rm raw}\rangle
\text{ with }|\mathcal I|=KM_{\rm fin},\nonumber\\
&\mathbb D
\left(Q_{\bm Z,\mathcal{C}}^{\mathcal I}
\middle\|Q_{\bm Z,0}^{(n)}\right)
\leq
G_n+\log\left(1+\frac{2^{\gamma_0}}{KM_{\rm fin}}\right)
\nonumber\\
&\hspace{2em}
+dn p_n^{(0)}+r_n-\log\Delta+\lambda_3
\Big\}
\end{align}
with $G_n=\mathbb D\left(Q_{\bm Z,1}^{(n)}\middle\|Q_{\bm Z,0}^{(n)}\right)$.
Then $\mathbb{P}(\mathcal{E}_1 \cap \mathcal{E}_2) > 0$ if 
\begin{equation}\label{capevents20}
\lambda_1^{-1}+\exp(-M_{\rm raw}q_n)<1
\end{equation}
with \begin{equation}
q_n:=
\frac{2\widehat\lambda_{2,n}\lambda_3^2}
     {n^2\Theta_n^2}
-(\ln2)\mathbb H_b(\widehat\lambda_{2,n}).
\end{equation}
and $\Theta_n$ in (\ref{Thetaexpression}).
\end{Lemma}
\noindent\emph{Proof:} See the Appendix.

\begin{Remark}
Different from Lemma 4 in \cite{M.Tahmasbi2}, here the event $\mathcal{E}_1$ only bounds the average error probabilities of the second type for all sub-codebooks. We have derived upper bound on the maximal decoding error probability of the first type in Lemma \ref{decodingerrortype1}.
\end{Remark} 
We have illustrated the random coding scheme, decoding method in Section \ref{oneshotachi}. The upper bound on the first type of decoding error in Lemma \ref{decodingerrortype1} and the average of the second type of decoding error has been analyzed in Lemma \ref{averagesecondtypeerror}. We also bounded the expectation of the discriminative metric - $\mathbb{D}(Q^{(n)}_{\bm{Z},\mathcal{C}}\| Q^{(n)}_{\bm{Z},0})$. A legitimate covert code should 
 be based on the previous Lemma \ref{existencesuff}. The choice of various parameters both reliability and covertness will be illustrated in our main results as next theorem to realize the event $\mathcal{E}_1 \cap \mathcal{E}_2$. We will provide characterizations on the size $M$ of each codebook for covert communication over AWGN channels with sufficient large $n$ through proper choice of the power level $\Psi(n)$. The theorem also characterizes the sufficient number of key bits $\log K$, and it is shown that $K$ depends not only on the noise level $\sigma^2$ at the adversary's channel and covert parameter $\delta$, but also depends on the hyper-parameter of truncation $\mu$. 
\begin{Theorem}\label{finiteachievabilty}
Let $M_{\rm raw}$ and $M_{\rm fin}$ be in (\ref{factM}).
The following conditions are sufficient for the
existence of an
\((M_{\rm fin},K,n,\epsilon,\sigma,\delta)\) code.
\begin{enumerate}
\item Reliability:
\begin{equation}\label{reliablity1}
\epsilon_1+
\frac{\lambda_1}{1-\widehat\lambda_{2,n}}
n^{-c_0}R_{Y,n}\leq\epsilon,
\end{equation}
where
\[
R_{Y,n}=
\mathbb E_{\bar Q_{\bm Y,1}^{(n)}}
\left[
\frac{\bar f_1^{(n)}(\bm Y)}
     {f_1^{(n)}(\bm Y)}
\right].
\]

\item Covertness:
\begin{equation}\label{covertness1}
\begin{aligned}
G_n
&+\log\left(1+\frac{2^{\gamma_0}}{KM_{\rm fin}}\right)\\
&+dn p_n^{(0)}+r_n-\log\Delta+\lambda_3
\leq\delta.
\end{aligned}
\end{equation}

\item Positive existence probability:
\begin{equation}\label{capevents3}
\lambda_1^{-1}
+\exp(-M_{\rm raw}q_n)<1,
\end{equation}
where
\[
q_n:=
\frac{2\widehat\lambda_{2,n}\lambda_3^2}
     {n^2\Theta_n^2}
-(\ln 2)\mathbb H_b(\widehat\lambda_{2,n}).
\]
\end{enumerate}
The choice of the related parameters to fulfill the above three conditions is listed as follows.  
\begin{enumerate}
\item [(a)] Choose intermediate parameters $\lambda_1,\lambda_2,\lambda_3$ as (1): $\lambda_1 = n^{c_1}$, (2) $\lambda_2 = 1 - n^{-c_2}$, (3): $\lambda_3 = n^{-c_3}$ such that $c_1,c_2,c_3$ all positive, meanwhile, they satisfy $0 < c_1 < c_0$, $c_1 + c_2 < c_0$ and $0 < 2c_3 < c_2 -2$. 
    
\item [(b)] \begin{equation}\label{logM_n1}
M_{\rm raw}:=
\left\lfloor2^{\gamma-c_0\log n}\right\rfloor,
\end{equation}
where \(\gamma\) is defined in
\eqref{gammadefintion}.
\item [(c)] 
Fix \(c_5>0\), and let \(\gamma_0\) be the
threshold defined in \eqref{gamma0definition}.
Set
\[
x_n:=\gamma_0+c_5\log n-\log M_{\rm raw},
\qquad
K_n:=\left\lceil2^{x_n}\right\rceil.
\]
Then
\begin{equation}\label{keynumbercondition}
\frac{2^{\gamma_0}}{M_{\rm raw}K_n}
\leq n^{-c_5}.
\end{equation}
Moreover,
\[
[x_n]^+\leq\log K_n<[x_n]^++1,
\qquad [x]^+:=\max\{0,x\}.
\]
If \(x_n\leq0\), then \(K_n=1\).
\item[(d)]
Choose \(\Psi(n)=c_4n^{-1/2}\), with \(c_4>0\),
so that the full covertness condition
\eqref{covertness1} holds.
In particular, the nominal Gaussian divergence
must satisfy
\[
G_n=
\mathbb D(Q_{\bm Z,1}^{(n)}
\|Q_{\bm Z,0}^{(n)})<\delta.
\]
\end{enumerate}
Furthermore, for sufficiently large $n$, we can choose $c_0 =13$, $c_1 = 1$, $c_2 = 11$, $c_3 = 4$, and $c_5 =3$ and a proper power level $c_4 = \frac{2\sigma^2}{\mu} \sqrt{\delta \ln 2}$. For the following asymptotic achievability and key-size statements, fix $\epsilon\in(0,1/2)$ and choose $\epsilon_1(n)=\epsilon-2/n$ for sufficiently large $n$. The uniform Berry--Esseen estimate in the supplementary proof gives $\beta_n=O(n^{-1/2})$, so $0<\epsilon_1(n)-\beta_n<1/2$ eventually. We obtain \begin{align}\label{Msizefinal0}
\log M^*(n,\epsilon,\sigma, \delta) & \geq  \mu \sigma^2 (\delta n\log e)^{1/2}  \nonumber\\
+  \sigma (\frac{2}{\mu})^{1/2}&  (n\delta)^{1/4} (\log e)^{3/4} \Phi^{-1}(\epsilon) + o(n^{\frac{1}{4}}).
\end{align}
For fixed $\mu\in(1/2,1)$ and $\epsilon\in(0,1/2)$, the
sufficient key size using $\gamma_0$ satisfies
\begin{equation}
\log K_n=
\begin{cases}
\bar{A}_{\mu}\sqrt{n}+Bn^{1/4}+o(n^{1/4}),
& \bar{A}_{\mu}>0,\\[3pt]
Bn^{1/4}+o(n^{1/4}),
& \bar{A}_{\mu}=0,\\[3pt]
0,
& \bar{A}_{\mu}<0,
\end{cases}
\end{equation}
where
\begin{align}
\bar{A}_{\mu}
&\triangleq
\frac{1-\mu^2\sigma^2}{\mu}\sqrt{\delta\log e},\\
B
&\triangleq
\sigma\sqrt{\frac{2}{\mu}}\,
\delta^{1/4}(\log e)^{3/4}\Phi^{-1}(1-\epsilon).
\end{align}
The three cases correspond to $\sigma^2<\mu^{-2}$,
$\sigma^2=\mu^{-2}$, and $\sigma^2>\mu^{-2}$,
respectively. 
\end{Theorem}
\noindent\emph{Proof:} See the Appendix.

\begin{Remark}[Truncated Gaussian random coding]
Unlike conventional Gaussian random coding
\cite[Theorem 5]{Ligong Wang},
\cite[Theorem 9.1.1]{Cover}, our construction
employs a truncated Gaussian distribution
supported on the truncated shell $\mathcal{A}_n$.
The upper truncation radius guarantees that
every codeword satisfies the prescribed power
constraint, while the lower radius excludes
low-energy codewords and facilitates a uniform
bound on the Type-I decoding error probability
(Lemma~\ref{decodingerrortype1}).
The underlying Gaussian distribution has variance
$\mu\Psi(n)$, so that $\mu$ determines both
the generating variance and the lower truncation
radius. This choice facilitates the reliability
and resolvability analyses, with the truncation
effect explicitly accounted for in the covertness
constraint. Although these parameters could be
selected independently, the present coupling
provides a tractable construction whose achievable
throughput coefficients depend on $\mu$.
\end{Remark}

\begin{Remark}[Covert power scaling]
Under the KL-divergence covertness constraint in
(\ref{covertconstraint}), the transmit power is chosen
to satisfy
$\Psi(n)=O\left(\frac{1}{\sqrt{n}}\right)$,
as specified in condition (d).
This scaling follows from the divergence between the
Gaussian output distributions
$Q_{\bm{Z},1}^{(n)}$ and $Q_{\bm{Z},0}^{(n)}$,
which determines the nominal covertness cost.
The same square-root power scaling can also arise under
other commonly used covertness measures, including
total variation distance, Hellinger distance, and
$\chi^2$-divergence. However, the admissible power
coefficient generally depends on the chosen divergence
measure and its corresponding covertness threshold.
Consequently, the achievable throughput coefficients
may differ, even when the scaling orders remain
unchanged. 
\end{Remark}

\begin{Remark}
Theorem~\ref{finiteachievabilty} extends the
achievability analysis in \cite[Theorem 5]{Ligong Wang}
in two aspects. First, it allows unequal noise
variances at the legitimate receiver and the
adversary, thereby revealing the effect of channel
asymmetry on the achievable throughput. In particular,
a larger adversarial noise variance permits higher
transmit power under the same covertness constraint
and consequently improves the achievable throughput.
Second, whereas \cite{Ligong Wang} characterizes
the leading-order throughput through an asymptotic
limit, our analysis provides explicit achievable
first- and second-order terms under a prescribed
decoding error probability $\epsilon$. These results
further quantify the finite-blocklength effects
associated with reliability, resolvability, and
truncated Gaussian coding.
\end{Remark}
\begin{Remark}[Interpretation of the key-size transition]
The regime-transition condition $\mu^2\sigma^2=1$
reflects a balance between the randomness supplied by
the uniformly distributed message index and the
randomness required by channel resolvability at Willie.
Reliability limits the message size $M_n$, whereas
resolvability requires a sufficiently large total
codebook size $M_nK_n$. The secret key supplies the
additional randomness when the message index alone
does not meet this requirement.

In the present analysis, the common decoding threshold
is controlled by the lower energy endpoint
$\mu^2n\Psi(n)$ of the truncation shell, while the
resolvability threshold $\gamma_0$ is chosen using
its upper endpoint $n\Psi(n)$. Their leading-order
difference is
\begin{equation}
\gamma_0-\log M_n
=
\frac{n\Psi(n)\log e}{2}
\left(\frac{1}{\sigma^2}-\mu^2\right)
+O(n^{1/4}).
\end{equation}
Consequently, when $\sigma^2<\mu^{-2}$, a positive
first-order randomness deficit remains, and the
derived sufficient key length scales as $\sqrt n$.
At $\sigma^2=\mu^{-2}$, the first-order terms balance,
but the second-order reliability penalty reduces
the message size, leaving a sufficient key length
of order $n^{1/4}$ for fixed $\epsilon\in(0,1/2)$.
When $\sigma^2>\mu^{-2}$, the message randomness
exceeds the resolvability requirement by a positive
first-order margin, and $K_n=1$ is sufficient for
all sufficiently large $n$.

Since Bob's noise variance is normalized to one,
a larger $\sigma^2$ means that Willie's channel is
relatively noisier, reducing the additional randomness
required from the key. However, Bob's channel being
better than Willie's, namely $\sigma^2>1$, does not
alone guarantee keyless operation under this bound:
the stronger condition $\sigma^2>\mu^{-2}$ is sufficient.
Increasing $\mu$ narrows the truncation shell and
reduces the separation between the energy endpoints
used in the analysis, lowering the threshold
$\mu^{-2}$ toward one. With general noise variances,
the corresponding transition is
$\sigma_w^2/\sigma_b^2=\mu^{-2}$.
These regimes describe sufficient key consumption
for the proposed construction; they do not establish
necessary key requirements for arbitrary covert codes.
\end{Remark}
Our results identify three contributions to the
covertness budget in \eqref{covertness1}. The Gaussian
reference term
$\mathbb D(Q_{\bm Z,1}^{(n)}\|Q_{\bm Z,0}^{(n)})$
captures the covertness cost associated with the
signal variance $\mu\Psi(n)$ relative to Willie's
noise variance $\sigma^2$. The resolvability and tail
penalties account for approximating the reference
output distribution with a finite codebook. Finally,
$r_n-\log\Delta$ accounts for restricting the Gaussian
input to a finite-energy shell. Thus, $\Psi(n)$ and
$\mu$ control the reference signal power and
truncation, while the codebook size $MK$ affects
the resolvability penalty.

The truncation is not necessary when we only consider the first order asymptotics, such as the proof of the achievability part of Theorem 5 in \cite{Ligong Wang}.
For a fixed truncation parameter \(\mu\in(\frac{1}{2},1)\), the achievable first-order throughput coefficient is \(\mu\sigma^2\sqrt{\delta\log e}\), which is strictly smaller than the benchmark coefficient \(\sigma^2\sqrt{\delta\log e}\) in Theorem \ref{conversetheorem}. Although choosing a blocklength-dependent parameter \(\mu_n\to1\) suggests a way to close this gap, such a choice cannot be justified by directly substituting \(\mu_n\) into the existing results. The reliability and resolvability estimates are established for fixed \(\mu\), and their constants and remainder terms have not been shown to remain controlled uniformly as \(\mu\to1\). 
 \subsection{Converse Bound}\label{Converse111}
In this section, the converse bound under covert constraint (\ref{covertconstraint}) is investigated, and the technique is to reconcile the bound in terms of the output distribution of the channel to get an upper bound of average power level of the codebook, which has been utilized in \cite{Hou2}\cite{Ligong Wang}. The difference is that the average power considered here should be over all messages and keys for $n$ channel uses. To prove the converse for any code corresponding to any key, we follow the argument in Theorem 7 of \cite{M.Tahmasbi2} and leverage the pigeon-hole principle in our arguments.

First, following the inequality (231) in \cite{M.Tahmasbi2}, we have
\begin{align}
\delta \geq \mathbb{D}(Q^{(n)}_{\bm{Z},\bm{c}}\| Q^{(n)}_{\bm{Z},0}) \geq n \mathbb{D}(\bar{Q}^{(n)}_{\bm{Z},\bm{c}}\| Q_{\bm{Z},0}),
\end{align}
where $\bar{Q}^{(n)}_{\bm{Z},\bm{c}} = \frac{1}{n}\sum_{i=1}^n\sum_{s=1}^{K}\sum_{w=1}^{M} \frac{1}{MK} W_2(\bm{z}(i)|\bm{x}_{sw}(i))$ is the average output distribution at the adversary per channel use over all the messages and keys and $Q_{\bm{Z},0}$ is the distribution of noise at the adversary. Following the same arguments around (74) in \cite{Ligong Wang},  we have the following 
lemma.
\begin{Lemma}\label{conversecorollary}
Fix \(\delta>0\) and \(\sigma>0\).
For any \((M,K,n,\epsilon,\sigma,\delta)\) code,
define its actual average power over all
messages and keys by
\[
\overline P_{\bm c}
:=
\frac{1}{nMK}
\sum_{s=1}^{K}\sum_{w=1}^{M}
\|\bm x_{sw}\|^2.
\]
If
\[
\mathbb D
\left(Q_{\bm Z,\bm{c}}^{(n)}
\middle\|Q_{\bm Z,0}^{(n)}\right)
\leq\delta,
\]
then
\[
\overline P_{\bm{c}}
\leq\Psi(\delta,\sigma,n),
\]
where \(\Psi(\delta,\sigma,n)\) is the unique
positive solution of
\begin{equation}\label{averagepowerlevel}
\frac n2
\left[
\frac{\Psi}{\sigma^2}
-\ln\left(1+\frac{\Psi}{\sigma^2}\right)
\right]\log e
=\delta.
\end{equation}
\end{Lemma}
\noindent\emph{Proof:}
See the Appendix.

 Lemma~\ref{conversecorollary} bounds the actual
average power over all messages and keys.
To apply a converse under a maximal-power
constraint, we next extract sufficiently many
low-energy codeword indices. 
\begin{Corollary}\label{conversecorollary2}
For \(n>1\), any
\((M,K,n,\epsilon,\sigma,\delta)\) code
has a set of message-key indices
\[
\mathcal D\subseteq
\langle K\rangle\times\langle M\rangle
\]
such that
\[
|\mathcal D|\geq
\left\lceil\frac{MK}{n}\right\rceil
\]
and
\[
\max_{(s,w)\in\mathcal D}
\frac{\|\bm x_{sw}\|^2}{n}
\leq
\frac{n}{n-1}\Psi(\delta,\sigma,n).
\]
\end{Corollary}
\noindent\emph{Proof:}
See the Appendix.

The above corollary shows that, for any $(M,K,n,\epsilon,\sigma,\delta)$ code code which satisfies the covert constraint,  there exists a non-negligible ``part" of the whole codewords such that the maximal power is bounded. By using existing converse bounds under maximal power constraint, we can upper bound the size of that part, and it leads to an upper bound of the size of each subcode. The result is illustrated in the following theorem. 
\begin{Theorem}\label{conversetheorem}
For covert communication over AWGN channels (\ref{channelmodel0})  and  (\ref{channelmodel1}), any  $(M,K,n,\epsilon, \sigma,\delta)$ code should satisfy
\begin{align}\label{conversefinal}
 \log  M^*(n, \epsilon,\sigma,\delta)
 &\leq \sigma^2 (n\delta \log e)^{1/2} \\
  + \sigma (2)^{1/2}& (n\delta)^{1/4}(\log e)^{3/4}\cdot  \Phi^{-1}(\epsilon)
  + O(\log n). \nonumber 
 \end{align} 
\end{Theorem}
\noindent\emph{Proof:} See the Appendix.

\begin{Remark}
The converse follows an approach similar to
\cite{M.Tahmasbi2}, which uses codeword-weight
bounds to derive a converse for covert binary-input
discrete memoryless channels. Here, KL divergence
facilitates the Gaussian optimality argument underlying
the converse. This argument does not extend directly
to total variation distance or other divergences.
\end{Remark}

\begin{Remark}
For noise variances $\sigma_b^2,\sigma_w^2>0$, let
\[
 \bm Y=\bm X+\bm N_b,\qquad
 \bm Z=\bm X+\bm N_w,
\]
where $\bm N_b\sim\mathcal N(\bm0,\sigma_b^2\bm I_n)$
and $\bm N_w\sim\mathcal N(\bm0,\sigma_w^2\bm I_n)$.
Dividing the input and both outputs by $\sigma_b$
yields the normalized model with
\[
 \sigma^2=\frac{\sigma_w^2}{\sigma_b^2},
 \qquad
 \Psi_{\mathrm{norm}}(n)
 =\frac{\Psi_{\mathrm{phys}}(n)}{\sigma_b^2}.
\]
This transformation preserves decoding error
probability and KL divergence, so our results
apply through these substitutions. The achievable
and converse first-order throughput coefficients
are, respectively,
\[
 \mu\frac{\sigma_w^2}{\sigma_b^2}
 \sqrt{\delta\log e}
 \quad\text{and}\quad
 \frac{\sigma_w^2}{\sigma_b^2}
 \sqrt{\delta\log e}.
\]
The second-order and sufficient key-size expressions
likewise depend on this noise-variance ratio.
Greater noise at Willie permits more signal power,
whereas greater noise at Bob reduces reliable
throughput. Their relative channel quality also
determines the sufficient key-size scaling.
\end{Remark}
\section{Numerical Results and Discussion}\label{sec:numerical}
We illustrate the asymptotic throughput achievability and converse expansions and sufficient secret-key scaling as functions of $n$, $\mu$, $\delta$, and $\sigma^2$. Unless specified otherwise, $\epsilon=10^{-3}$ and $10^3\le n\le10^5$. The curves are evaluated from the retained first- and second-order terms, omitting higher-order remainders. They illustrate asymptotic trends rather than certified finite-blocklength bounds.

\subsection{Truncation Parameter and Noise Variance}
\begin{figure*}[t]
\centering
\includegraphics[width=0.96\textwidth]{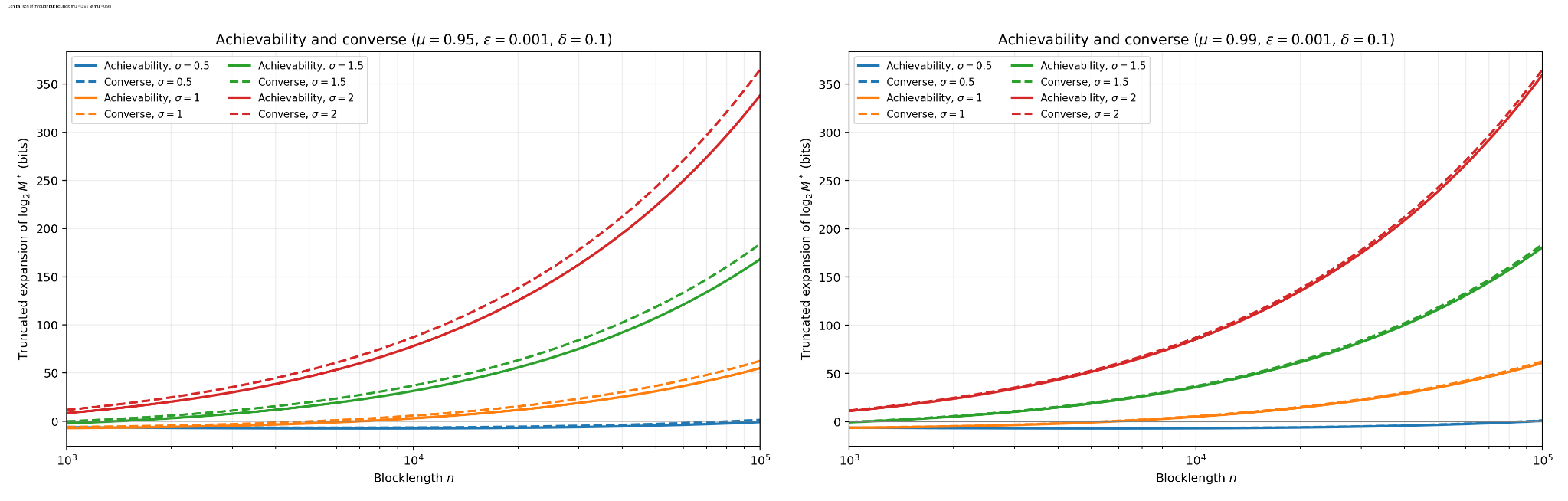}
\caption{Truncated throughput achievability (solid)
and converse (dashed) expansions for
$\mu=0.95,0.99$, $\delta=0.1$, and different
$\sigma$ ($\epsilon=10^{-3}$).}
\label{fig:mu_comparison}
\end{figure*}

Fig.~\ref{fig:mu_comparison} compares
$\mu\in\{0.95,0.99\}$ for $\delta=0.1$ and
$\sigma\in\{0.5,1,1.5,2\}$. The leading terms
predict increasing throughput with blocklength
and adversary noise variance. At smaller
blocklengths, however, the negative second-order
correction can produce negative or nonmonotone
approximations that provide no useful positive
throughput estimate. The achievable and converse
first-order coefficients are
$\mu\sigma^2\sqrt{\delta\log e}$ and
$\sigma^2\sqrt{\delta\log e}$, respectively.
Their difference,
$(1-\mu)\sigma^2\sqrt{\delta\log e}$,
decreases as $\mu$ approaches one. Thus,
$\mu=0.99$ yields a smaller gap than $\mu=0.95$,
although the coefficients remain unequal for
fixed $\mu<1$. The effect of $\sigma^2$ becomes
more pronounced at larger $n$, consistent with
the leading $\sqrt n$ contribution.

\subsection{Covertness Budget and Throughput--Key Tradeoff}
\begin{figure*}[t]
\centering
\includegraphics[width=0.96\textwidth]{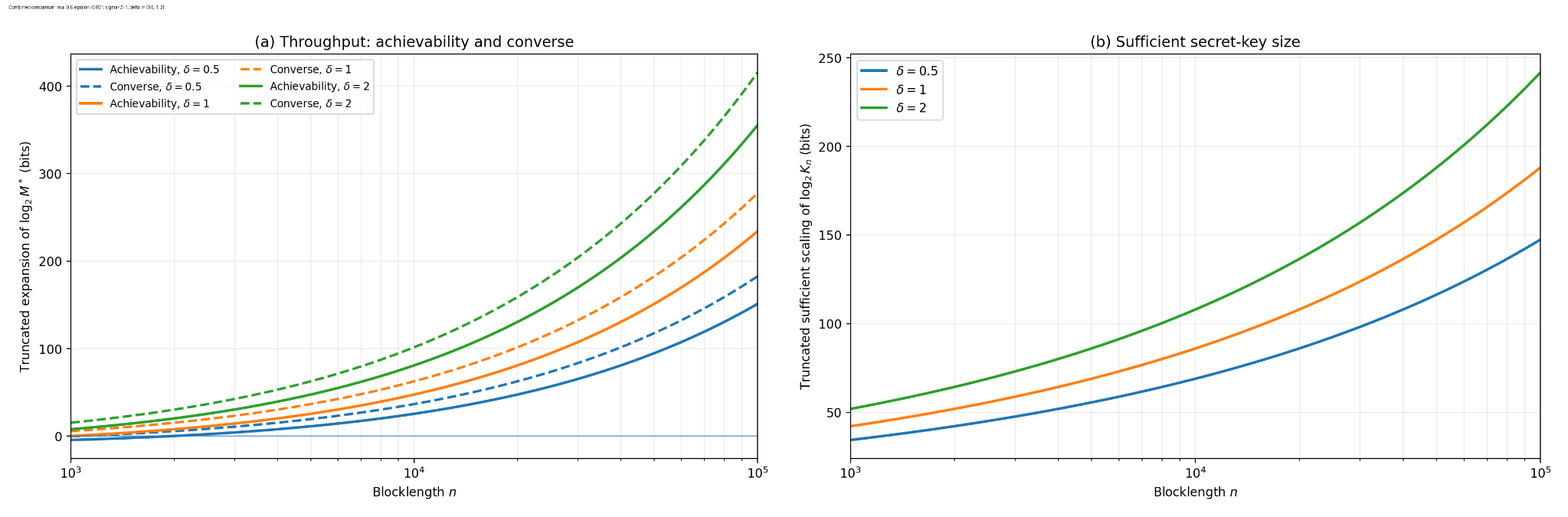}
\caption{For $\mu=0.9$, $\sigma^2=1$, and
$\delta\in\{0.5,1,2\}$: (a) truncated throughput
expansions; (b) sufficient secret-key expansions
($\epsilon=10^{-3}$).}
\label{fig:throughput_key}
\end{figure*}

Fig.~\ref{fig:throughput_key}(a) compares throughput
expansions for $\mu=0.9$, $\sigma^2=1$, and
$\delta\in\{0.5,1,2\}$. Increasing $\delta$ relaxes
covertness and raises both approximations, whose
leading coefficients scale as $\sqrt\delta$.
For fixed $\mu$, the absolute first-order gap
also grows as $\sqrt\delta$. The $n^{1/4}$
corrections noticeably affect moderate-blocklength
behavior, which leading-order asymptotics alone
may not capture.

Fig.~\ref{fig:throughput_key}(b) shows the
corresponding sufficient secret-key scaling.
For these parameters, a larger covertness budget
increases both achievable throughput and the
sufficient key-size expression, which contains
a positive $\sqrt n$ term and an $n^{1/4}$
correction. This illustrates a throughput--key
consumption tradeoff for our construction, rather
than a key-size lower bound for arbitrary schemes.

\subsection{Noise-Dependent Secret-Key Scaling}
\begin{figure}[t]
\centering
\includegraphics[width=\columnwidth]{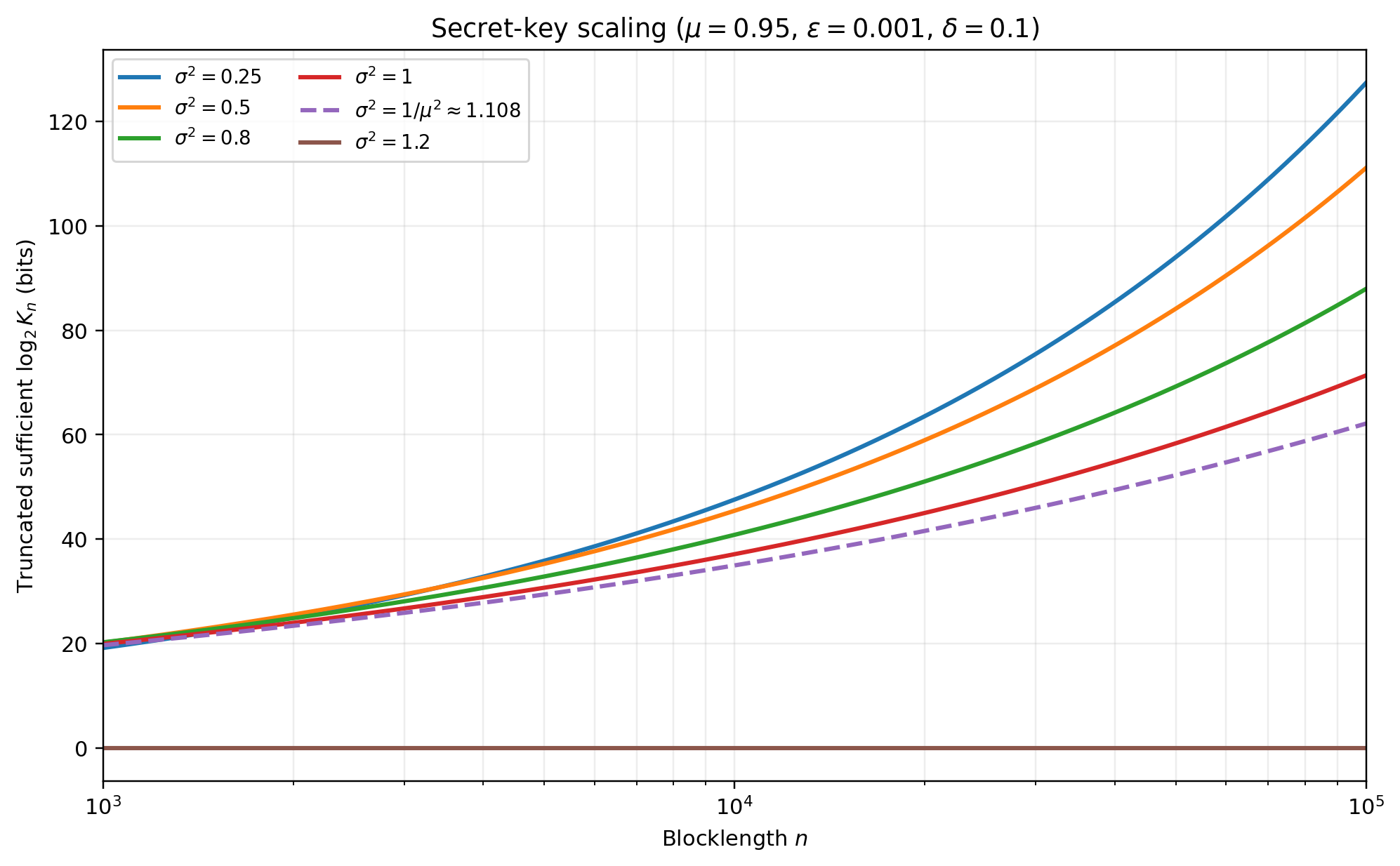}
\caption{Sufficient key-size expansions for
$\mu=0.95$, $\delta=0.1$, and different $\sigma^2$;
$\mu^{-2}\approx1.108$ ($\epsilon=10^{-3}$).}
\label{fig:key_noise}
\end{figure}

Fig.~\ref{fig:key_noise} shows the key-size
expansion for $\mu=0.95$ and $\delta=0.1$.
The threshold $\sigma^2=\mu^{-2}\approx1.108$
separates three regimes. Below it,
$\bar A_\mu>0$ gives a leading sufficient
key-size term of order $\sqrt n$. At equality,
this term vanishes, leaving an $n^{1/4}$
contribution when its coefficient is positive.
Above the threshold, the construction permits
$K_n=1$ for sufficiently large $n$.
The zero-key curve therefore represents an
asymptotic sufficiency result for this construction.
Greater adversary noise can thus reduce sufficient
key consumption and change its asymptotic order.

Overall, the plots illustrate square-root throughput
scaling, the fixed-$\mu$ achievability--converse gap,
and the noise-dependent key threshold. Second-order
terms affect moderate-blocklength behavior.
Establishing a matching first-order coefficient
through $\mu_n\to1$ requires uniform reliability,
resolvability, and expurgation estimates.
\section{Conclusion}
This paper studies finite-blocklength covert
communication over AWGN channels under a
KL-divergence constraint. Using truncated Gaussian
random coding, resolvability, and expurgation,
we derive explicit achievable first- and second-order
throughput terms and sufficient secret-key scaling.
A converse follows by translating covertness into
an average-power bound and extracting a subcode
with controlled maximal power. These results
quantify the effects of blocklength, decoding error,
adversary noise, and covertness budget.

For fixed $\mu<1$, the first-order
achievability--converse gap arises from the
construction and analysis, rather than an established
fundamental limitation. Likewise, the key-size
bound is sufficient, not a minimum over all codes.
Allowing $\mu_n\to1$ may close the throughput gap,
but requires uniform reliability, resolvability,
and expurgation estimates with controlled residual
covertness penalties. Matching second-order bounds and necessary secret-key requirements remain directions for future work.
\appendix
\subsection{Proof of Lemma 1}\label{proofofLemma1}
\begin{proof}
From Definition 3, an $(M,K,n,\epsilon,\Psi,\sigma, \delta)_m$ is necessarily an $(M,n,\epsilon,\Psi)_m$ code, hence the left-hand of (16) holds.  An $(M,K,n,\epsilon,\Psi,\sigma, \delta)_m$ is necessarily an $(M,K,n,\epsilon,\sigma,\delta)$ code, and an $(M,K,n,\epsilon,\Psi,\sigma, \delta)_a$ is necessarily a $(M,K,n,\epsilon,\sigma,\delta)$ code,  hence, we have the inequalities (17) and the right-hand of (18). The inequalities on right-hand of (16) and the left-hand of (18) follow from the fact that a code which satisfies maximal power constraint less or equal than $\Psi$ necessarily has average power less or equal than $\Psi$. 
\end{proof}
\subsection{Proof of Lemma 2}\label{KLTVDcomputation101}
\begin{proof}
Let $\mathcal{A}_n
=\left\{\bm{x}:
\mu\sqrt{n\Psi(n)}
\leq\|\bm{x}\|
\leq\sqrt{n\Psi(n)}\right\}$, from the definition of $\bar{P}_n^n$ and  $P_n^n$, we have $\frac{\bar{g}^{(n)}(\bm{x})}{g^{(n)}(\bm{x})} = \frac{1}{\Delta}$ for $\bm{x} \in \mathcal{A}_n$, and $\bar{g}^{(n)}(\bm{x}) = 0$ for $\bm{x} \in \mathcal{A}_n^c$. Hence, we have
\begin{equation}
\begin{split}
\mathbb{D}(\bar{P}_n^{(n)}, P_n^{(n)}) = & \int_{\bm{x} \in \mathbb{R}^n} \bar{g}^{(n)}(\bm{x}) \log \frac{\bar{g}^{(n)}(\bm{x})}{g^{(n)}(\bm{x})} d\bm{x}\\
= & \int_{\bm{x} \in \mathcal{A}_n} \bar{g}^{(n)}(\bm{x}) \log \frac{1}{\Delta} d\bm{x}\\
= & -\log \Delta.
\end{split}
\end{equation}
and
\begin{align}\label{trunction}
\mathbb{V}(\bar{P}_n^{(n)}, P_n^{(n)})= & \frac{1}{2}\underset{\mathbb{R}^n}{\int}|\bar{g}^{(n)}(\bm{x}) - g^{(n)}(\bm{x})|d\bm{x} \nonumber\\
= & \frac{1}{2} \big[\int_{\bm{x}\in \mathcal{A}_n} (\frac{1}{\Delta} -1 )g^{(n)}(\bm{x}) d\bm{x} \nonumber \\
 + & \int_{\bm{x}\in \mathcal{A}_n^c} g^{(n)}(\bm{x})d\bm{x}\big] \nonumber\\
= & \frac{1}{2}\big[ (\frac{1}{\Delta} -1 ) \Delta + 1 -\Delta\big] \nonumber\\
= & 1 -\Delta.
\end{align}
The proof of equation (\ref{constantmu}) of the main text can be found in \cite{Yu1}.
\end{proof}
\subsection{Proof of Lemma 3}
\begin{proof}
By Bayes' formula,
\begin{align}
\frac{\bar f_1^{(n)}(\bm y)}{f_1^{(n)}(\bm y)}
= \frac{\mathbb{P}[\bm{x} \in \mathcal{A}_n | \bm{y} ]}{\mathbb{P}[\bm{x} \in \mathcal{A}_n]}
=\frac{\mathbb{P}\{\bm X\in\mathcal{A}_n\mid\bm Y=\bm y\}}\Delta
\leq\frac1\Delta.
\end{align}
Positivity follows from the positive Gaussian kernel and $\Delta>0$.
The same argument applies to the ratio $\frac{\bar h_1^{(n)}(\bm z)}{h_1^{(n)}(\bm z)}$. The two inequalities in (35) follow from applying data processing inequalities for the conclusion of Lemma 2.
\end{proof}
\subsection{Proof of Lemma 4}
\begin{proof}
Fix a codeword $\bm{x}_{sw}$ in the support of
$\bar P_n^{(n)}$, and write
$\|\bm{x}_{sw}\|^2=n\Upsilon$, where
$\Upsilon\in[\mu^2\Psi(n),\Psi(n)]$.
By the rotational symmetry of the AWGN channel and the
reference distribution $Q_{\bm Y,1}^{(n)}$, the information
density under $W_1^n(\cdot|\bm{x}_{sw})$ depends on the
codeword only through its energy. We may therefore take
\[
 \bm{x}_0=(\sqrt{\Upsilon},\ldots,\sqrt{\Upsilon}).
\]
For this codeword, the information density is
\begin{align}
 \log\frac{W_1^n(\bm Y|\bm{x}_0)}
 {f_1^{(n)}(\bm Y)}
 &=
 \frac n2\log(1+\mu\Psi(n))
 \nonumber\\
 &\quad+
 \frac{\log e}{2}\sum_{i=1}^n
 \left[
 \frac{Y_i^2}{1+\mu\Psi(n)}
 -(Y_i-\sqrt{\Upsilon})^2
 \right].
\end{align}
Under $W_1^n(\cdot|\bm{x}_0)$, we have
$Y_i=\sqrt{\Upsilon}+\xi_i$, where
$\xi_1,\ldots,\xi_n$ are independent standard normal
random variables. Hence, the information density has
the same distribution as
\begin{align}
 H_n^\Upsilon
 &=
 \frac n2\log(1+\mu\Psi(n))
 \nonumber\\
 &\quad+
 \frac{\log e}{2(1+\mu\Psi(n))}
 \sum_{i=1}^n
 \left[
 \Upsilon-\mu\Psi(n)\xi_i^2
 +2\sqrt{\Upsilon}\xi_i
 \right].
 \label{H_n}
\end{align}
Write
\[
 H_n^\Upsilon
 =n\mathsf C(\mu,\Psi(n))
 +\sum_{i=1}^nS_i(\Upsilon,\mu\Psi(n)),
\]
where
\[
 S_i(\Upsilon,\mu\Psi(n))
 :=
 \frac{\log e}{2(1+\mu\Psi(n))}
 \left[
 \Upsilon-\mu\Psi(n)\xi_i^2
 +2\sqrt{\Upsilon}\xi_i
 \right].
\]
For each fixed $n$ and $\Upsilon$, these summands are
independent and identically distributed, with mean
\begin{equation}\label{capacityadditional1}
 \bar{\mathsf C}(\Upsilon,\mu\Psi(n))
 :=
 \mathbb E S_i
 =
 \frac{\Upsilon-\mu\Psi(n)}
 {2(1+\mu\Psi(n))}\log e
\end{equation}
and variance
\begin{equation}\label{Vdefinition1}
 \mathsf V(\Upsilon,\mu\Psi(n))
 :=
 \mathsf{Var}(S_i)
 =
 \frac{2\Upsilon+\mu^2\Psi(n)^2}
 {2(1+\mu\Psi(n))^2}\log^2 e.
\end{equation}
Both quantities are increasing in $\Upsilon$.
In particular,
\begin{align}
 \bar{\mathsf C}(\Upsilon,\mu\Psi(n))
 &\geq
 \bar{\mathsf C}(\mu^2\Psi(n),\mu\Psi(n)),
 \label{Ccompare1}\\
 \frac{\mathsf V(\Psi(n),\mu\Psi(n))}
 {\mathsf V(\Upsilon,\mu\Psi(n))}
 &=
 \frac{2\Psi(n)+\mu^2\Psi(n)^2}
 {2\Upsilon+\mu^2\Psi(n)^2}
 \geq1.
 \label{Vratio}
\end{align}

We next bound the Berry--Esseen coefficient uniformly
over $n$ and $\Upsilon$. Let
\[
 a_n:=\mu\Psi(n),\qquad
 b_\Upsilon:=2\sqrt{\Upsilon},\qquad
 c_n:=\frac{\log e}{2(1+\mu\Psi(n))}.
\]
Then
\[
 S_i-\mathbb E S_i
 =
 c_n[-a_n(\xi_i^2-1)+b_\Upsilon\xi_i].
\]
Using
$\mathbb E[(\xi_i^2-1)\xi_i]=0$, we obtain
\[
 \mathsf V(\Upsilon,\mu\Psi(n))
 =c_n^2(2a_n^2+b_\Upsilon^2).
\]
Define the finite constants
\[
 m_\chi:=\mathbb E|\xi_i^2-1|^3,
 \qquad
 m_G:=\mathbb E|\xi_i|^3.
\]
The inequality $|x+y|^3\leq4(|x|^3+|y|^3)$ yields
\[
 \mathsf T[S_i]
 :=\mathbb E|S_i-\mathbb E S_i|^3
 \leq4c_n^3(a_n^3m_\chi+b_\Upsilon^3m_G).
\]
Consequently, the Berry--Esseen coefficient
\[
 B(\Upsilon,\mu\Psi(n))
 :=
 \frac{6\mathsf T[S_i]}
 {\mathsf V(\Upsilon,\mu\Psi(n))^{3/2}}
\]
satisfies
\begin{align}
 B(\Upsilon,\mu\Psi(n))
 &\leq24
 \frac{a_n^3m_\chi+b_\Upsilon^3m_G}
 {(2a_n^2+b_\Upsilon^2)^{3/2}}
 \nonumber\\
 &\leq24\left(\frac{m_\chi}{2^{3/2}}+m_G\right)
 =:B_*.
\end{align}
The last inequality follows by bounding the denominator
separately by $2^{3/2}a_n^3$ and $b_\Upsilon^3$.
Thus, $B_*$ is independent of $n$ and $\Upsilon$.
Defining
\[
 \beta_n:=
 \frac1{\sqrt n}
 \sup_{\Upsilon\in[\mu^2\Psi(n),\Psi(n)]}
 B(\Upsilon,\mu\Psi(n)),
\]
we have
\[
 0\leq\beta_n\leq\frac{B_*}{\sqrt n}.
\]

Let
\begin{align*}
 m_n(\Upsilon)
 &:=
 n\mathsf C(\mu,\Psi(n))
 +n\bar{\mathsf C}(\Upsilon,\mu\Psi(n)),\\
 v_n(\Upsilon)
 &:=
 \sqrt{n\mathsf V(\Upsilon,\mu\Psi(n))},\\
 z_n&:=\epsilon_1-\beta_n.
\end{align*}
For fixed $0<\epsilon_1<1/2$, we have
$z_n\in(0,1/2)$ for all sufficiently large $n$.
Choose the decoding threshold such that
\begin{equation}\label{uniformthreshold}
 \gamma\leq
 m_n(\mu^2\Psi(n))
 +v_n(\Psi(n))\Phi^{-1}(z_n).
\end{equation}
By \eqref{Ccompare1}, this choice ensures that
\[
 \gamma-m_n(\Upsilon)
 \leq v_n(\Psi(n))\Phi^{-1}(z_n)
\]
for every $\Upsilon$ in the shell.
The Berry--Esseen inequality therefore gives
\begin{align}
 \mathbb{P}\{H_n^\Upsilon\leq\gamma\}
 &\leq
 \Phi\!\left(
 \frac{\gamma-m_n(\Upsilon)}{v_n(\Upsilon)}
 \right)+\beta_n
 \nonumber\\
 &\leq
 \Phi\!\left(
 \frac{v_n(\Psi(n))}{v_n(\Upsilon)}
 \Phi^{-1}(z_n)
 \right)+\beta_n
 \nonumber\\
 &\leq z_n+\beta_n
 =\epsilon_1.
\end{align}
Here, the final inequality follows from
\eqref{Vratio} and $\Phi^{-1}(z_n)<0$.

This bound holds uniformly over
$\Upsilon\in[\mu^2\Psi(n),\Psi(n)]$.
Since every codeword generated according to
$\bar P_n^{(n)}$ lies in this shell, rotational symmetry
and the preceding bound imply, for all sufficiently
large $n$,
\begin{equation}\label{maximalerrortype1}
 \max_{s,w}\varepsilon_{s,w}^{(1)}
 \leq\epsilon_1.
\end{equation}
\end{proof}

\subsection{Proof of Lemma 7}
\begin{proof}
By exchangeability, condition on a selected codeword $\bm X$ and
its channel output $\bm Z$. Conditional Jensen's inequality applied
to the other $MK-1$ codewords gives
\begin{align*}
&\mathbb{E}_{\mathcal{C}}\left[\mathbb{D}(Q^{(n)}_{\bm{Z},\mathcal{C}}\| Q^{(n)}_{\bm{Z},1})\right] \\
&\overset{(a)}{\leq}\mathbb E_{\bar P_n^{(n)}W_2^n}
\log\left[
\frac{W_2^n(\bm Z\mid\bm X)}{MK\,Q_{Z,1}^{(n)}(\bm Z)}
+\frac{MK-1}{MK}\frac{\bar Q_{Z,1}^{(n)}(\bm Z)}{Q_{Z,1}^{(n)}(\bm Z)}
\right]\\
&\leq\mathbb E_{\bar P_n^{(n)}W_2^n}
\log\left[\frac{W_2^n(\bm Z\mid\bm X)}{MK\,Q_{Z,1}^{(n)}(\bm Z)}
+\frac1\Delta\right]\\
&\leq-\log\Delta+\mathbb E_{\bar P_n^{(n)}W_2^n}
\log\left[1+\frac{W_2^n(\bm Z\mid\bm X)}{MK\,Q_{Z,1}^{(n)}(\bm Z)}\right],
\end{align*}
where we have used the fact that $\bm{X}_{sw}$ are i.i.d samples from $\bar{P}_n^{(n)}$ in $(a)$ and the last step follows $0 < \Delta < 1$.
\end{proof}
\subsection{Proof of Lemma 8}\label{WvsQ1}
\begin{proof}
\begin{align}
& \frac{W_2^n(\bm{z}|\bm{x})}{Q_{\bm Z,1}^{(n)}(\bm z)} \nonumber\\
 =  & \frac{\frac{1}{(2\pi\sigma^2)^{\frac{n}{2}}} e^{-\frac{\|\bm{z} -\bm{x}\|^2}{2\sigma^2}}}{\frac{1}{\big(2\pi(\sigma^2+ \mu \Psi(n))\big)^{\frac{n}{2}}} e^{-\frac{\|\bm{z}\|^2}{2(\sigma^2+ \mu \Psi(n))}}}\nonumber\\
 =  & \bigg(\frac{\sigma^2 + \mu \Psi(n)}{\sigma^2}\bigg)^{\frac{n}{2}} \nonumber \\
 \times &e^{\frac{-\mu \Psi(n)\|\bm{z}\|^2 + 2(\sigma^2 + \mu \Psi(n)) \bm{z}^{T}\bm{x} - (\sigma^2 + \mu \Psi(n))\|\bm{x}\|^2 }{2\sigma^2(\sigma^2 + \mu \Psi(n))}}\nonumber\\
  =  & \bigg(\frac{\sigma^2 + \mu \Psi(n)}{\sigma^2}\bigg)^{\frac{n}{2}} e^{-\frac{\mu\Psi(n)\big \|\bm{z}- \frac{\sigma^2 + \mu \Psi(n) }{\mu \Psi(n)}\bm{x}\big\|^2 }{2\sigma^2(\sigma^2 + \mu \Psi(n))} +  \frac{\|\bm{x}\|^2}{2\mu \Psi(n)}}\nonumber\\
\leq & \bigg(\frac{\sigma^2 + \mu \Psi(n)}{\sigma^2}\bigg)^{\frac{n}{2}} e^{ \frac{\|\bm{x}\|^2}{2\mu \Psi(n)}}\nonumber\\
\leq & \bigg(\frac{\sigma^2 + \mu \Psi(n)}{\sigma^2}\bigg)^{\frac{n}{2}} e^{ \frac{n}{2\mu }}.
\end{align}
\end{proof}
\subsection{Proof of Lemma 9}
\begin{proof}
The term $\mathbb{P}(\mathcal{E}^c)$ can be rewritten as follows
\begin{equation}\label{resolvabiltyfirsterm}
\begin{split}
 &\mathbb{P}(\mathcal{E}^c)
 =  \mathbb{P}_{\bar{P}_n^{(n)}(\bm{X})W^n_2(\bm{Z}|\bm{X})}\bigg(\log \frac{W_2^n(\bm{Z}|\bm{X})}{{Q}^{(n)}_{\bm{Z},1}(\bm{Z})} > \gamma_0\bigg) \\
 = & \mathbb{P}_{\bar{P}_n^{(n)}(\bm{X})}  \left[ \mathbb{P}_{W_2^n(\bm{Z}|\bm{X} = \bm{x})} \bigg(  \log \frac{W_2^n(\bm{Z}|{\bm{X} =\bm{x}})}{h_1^{(n)}(\bm{Z})}>\gamma_0  \bigg) \right].
\end{split}
\end{equation} 
Note that in (\ref{resolvabiltyfirsterm}), the distribution ${Q}^{(n)}_{\bm{Z},1}$ is the induced distribution of the channel $W_2^n$ by $P_n^{(n)}$ but not $\bar{P}_n^{(n)}$, which is an isotropic Gaussian distribution. The inner term in (\ref{resolvabiltyfirsterm})
\begin{equation}\label{innerprobability}
\mathbb{P}_{W_2^n(\bm{Z}|\bm{X} = \bm{x})} \bigg(  \log \frac{W_2^n(\bm{Z}|{\bm{X} =\bm{x}})}{h_1^{(n)}(\bm{Z})}> \gamma_0  \bigg)
\end{equation} 
is considered for each $\bm{x}_{sw}$ in the support set of $\bar{P}_n^{(n)}$. 
 Similarly as the proof of Lemma 4, we first consider the probability of error in (\ref{innerprobability}) for $\bm{x}_{sw}$ on the $n-1$-sphere with radius $\sqrt{n\Upsilon}$. The sphere symmetry allows us to assume  $\bm{x}_0 = [\sqrt{\Upsilon}, \cdots, \sqrt{\Upsilon}]$. we assume a codeword $\bm{x}_0 = [\sqrt{\Upsilon}, \cdots, \sqrt{\Upsilon}]$, where $\Upsilon \in [\mu^2\Psi,\Psi] $. The information density $\log \frac{W_2^n(\bm{z}|{\bm{x}_0})}{h_1^{(n)}(\bm{z})} $ is expressed as 
\begin{align}\label{informationdensity2}
& \frac{n}{2}\log \big(1 + \frac{\mu \Psi(n)}{\sigma^2}\big) \\
+  &\frac{\log e}{2}\sum_{i=1}^n \left[\frac{z_i^2}{\sigma^2 + \mu \Psi(n)} - \frac{(z_i -\sqrt{\Upsilon})^2}{\sigma^2}\right] \nonumber\\
= & \frac{n}{2}\log (1 + \frac{\mu \Psi(n)}{\sigma^2}) - \sum_{i=1}^n\frac{\log e}{2\sigma^2 (\sigma^2 + \mu \Psi(n))} \nonumber\\
 \times & \left[ \mu \Psi(n) z_i^2 -2 (\sigma^2 + \mu \Psi(n)) \sqrt{\Upsilon} z_i + (\sigma^2 + \mu \Psi(n))\Upsilon \right] \nonumber
\end{align}
 Under $W_2^n(\bm{Z}|{\bm{x}_0})$, we have $z_i = \sqrt{\Upsilon} + \sigma \xi_i$, then (\ref{informationdensity2})  has the same distribution as 
\begin{align}\label{barH_n}
I_n^\Upsilon =  &\frac{n}{2}\log \big(1 + \frac{\mu \Psi(n)}{\sigma^2}\big) \nonumber\\ + &\frac{\log e}{2\sigma^2(\sigma^2 + \mu \Psi(n))} 
\sum_{i=1}^n \bigg(\sigma^2\Upsilon - \mu \Psi(n)\sigma^2\xi_i^2 \nonumber\\
 + &  2\sigma^3\sqrt{\Upsilon}\xi_i\bigg)
\end{align}
where we have used $\xi_i \sim \mathcal{N}(0,1)$ with $i= 1,\cdots,n$.
We rewrite (\ref{barH_n}) as
\begin{align}
I_n^\Upsilon = &n\widetilde{\mathsf C}(\mu,\Psi(n),\sigma^2) \\
+ & \frac{\log e}{2(\sigma^2+\mu \Psi(n))}
\left[n\Upsilon-\mu \Psi(n) T_n+2\sigma\sqrt\Upsilon L_n\right]\nonumber
\end{align}
with
\begin{align}
T_n={}&\sum_{i=1}^n\xi_i^2,\qquad L_n=\sum_{i=1}^n\xi_i.
\end{align}
Define $A_n:=\frac{\log e}{2(\sigma^2+\mu\Psi(n))}$
and the expectation of $I_n^\Upsilon$ as a function of $\Upsilon$ is
\begin{equation}
 m_n^\Upsilon:=\mathbb{E} I_n^\Upsilon
 =n\widetilde{\mathsf{C}}_n+A_nn[\Upsilon-\mu\Psi(n)]
\end{equation}
where $\widetilde{\mathsf C}_n
=\widetilde{\mathsf C}(\mu,\Psi(n),\sigma^2)$.
We subtract $I_n^\Upsilon$ and $\gamma_0$ with the mean to get
\begin{align}
 I_n^\Upsilon-m_n^\Upsilon
 &=A_n[-\mu\Psi(n)(T_n-n)+2\sigma\sqrt\Upsilon L_n],\label{eq:center}\\
 \gamma_0-m_n^\Upsilon&=A_nn[\Psi(n)-\Upsilon].\label{eq:gap}
\end{align}
The probability in (\ref{resolvabiltyfirsterm}) can be reformulated as 
\begin{align}
&\mathbb{P}(\mathcal{E}^c) \nonumber\\
 = & \mathbb{P}_{\bar{P}_n^{(n)}(\bm{X})}  \left[ \mathbb{P}_{W_2^n(\bm{Z}|\bm{X} = \bm{x})} \bigg(  \log \frac{W_2^n(\bm{Z}|{\bm{X} =\bm{x}})}{h_1^{(n)}(\bm{Z})}>\gamma_0  \bigg) \right] \nonumber\\
 = &\mathbb{E}_{\Upsilon} [\mathbb{P}(I_n^\Upsilon>\gamma_0 \mid \Upsilon)]\nonumber\\
 =&\mathbb{P}_{\overline P_n^{(n)}} [\mathbb{P}(I_n^\Upsilon>\gamma_0\mid \Upsilon)]
\end{align}
\subsubsection{Bound for each $\Upsilon \in [\mu^2\Psi(n),\Psi(n))$}
Assume $\mu^2\Psi(n)\leq\Upsilon<\Psi(n)$, which excludes $\Upsilon = \Psi(n)$ and let
\[
 d_n(\Upsilon)=n[\Psi(n)-\Upsilon]>0,
 \qquad u_\Upsilon=\frac{\Psi(n)-\Upsilon}{2\mu\Psi(n)} > 0.
\]
Equations~\eqref{eq:center}--\eqref{eq:gap} and $A_n>0$ imply the exact event identity
\[
 \{I_n^\Upsilon>\gamma_0\}
 =\{-\mu\Psi(n)(T_n-n)+2\sigma\sqrt\Upsilon L_n>d_n(\Upsilon)\}.
\]
If a sum exceeds $d_n(\Upsilon)$, at least one summand exceeds $d_n(\Upsilon)/2$. Therefore
\begin{align}\label{eq:split}
 \mathbb{P}(I_n^\Upsilon>\gamma_0)
 &\leq\mathbb{P}\left(-\mu\Psi(n)(T_n-n)>\frac{d_n(\Upsilon)}2\right)\notag\\
 &\quad+\mathbb{P}\left(2\sigma\sqrt\Upsilon L_n>\frac{d_n(\Upsilon)}2\right).
\end{align}
Note that the above union bound does not depend on the independence of $T_n$ and $L_n$. We consider the bounding of (\ref{eq:split}) in the following steps: (a) (b) and (c). 
\begin{enumerate}
\item [(a)] Bound the first probability

The first event in~\eqref{eq:split} is $T_n<n(1-u_\Upsilon)$.  As $\Upsilon \in [\mu^2\Psi,\Psi)$,
\[
 0<u_\Upsilon\leq\frac{1-\mu^2}{2\mu}<1.
\]
For any $t>0$, we apply Markov's inequality to $e^{-tT_n}$ to get
\begin{align} \label{inequalityboud111}
 & \mathbb{P}(T_n<n(1-u_\Upsilon)) \\
 &\leq e^{tn(1-u_\Upsilon)}E e^{-tT_n} \nonumber\\
 &=\exp\left\{n\left[t(1-u_\Upsilon)-\frac12\ln(1+2t)\right]\right\},\nonumber
\end{align}
where $E e^{-tT_n}=(1+2t)^{-n/2}$. The derivative of the exponent per coordinate is $(1-u_\Upsilon)-(1+2t)^{-1}$. Its second derivative is positive, so the minimizing parameter $t$ for (\ref{inequalityboud111}) is
\[
 t_* =\frac{u_\Upsilon}{2(1-u_\Upsilon)}.
\]
We substitute $t_*$ to (\ref{inequalityboud111}) to have
\begin{align}
\mathbb{P}(T_n<n(1-u_\Upsilon))\leq
 \exp\left[-\frac n2\{-u_\Upsilon-\ln(1-u_\Upsilon)\}\right]. \nonumber
\end{align}
Also, 
\begin{align}\label{tempbound}
 -u_\Upsilon-\ln(1-u_\Upsilon)
 &=\int_0^{u_\Upsilon}
   \left(\frac{1}{1-v}-1\right)\,dv \nonumber\\
 &=\int_0^{u_\Upsilon}\frac{v}{1-v}\,dv \nonumber\\
 &\geq\int_0^{u_\Upsilon}v\,dv\nonumber\\
 &=\frac{u_\Upsilon^2}{2}.
\end{align}
Thus the first probability is at most $e^{-nu_\Upsilon^2/4}$.

\item [(b)] Bound the second probability

The second event is $L_n>s_\Upsilon$, with
$s_\Upsilon=\frac{d_n(\Upsilon)}{4\sigma\sqrt\Upsilon} $.
Since $E e^{tL_n}=e^{nt^2/2}$, for $t>0$, we have 
\begin{align}
 \mathbb{P}(L_n>s_\Upsilon)\leq e^{-ts_\Upsilon+nt^2/2},
 \end{align}
 by Chernoff bounding. 
The minimizing parameter is $t_*=s_\Upsilon/n$, and therefore
\[
 \mathbb{P}(L_n>s_\Upsilon)
 \leq\exp\left[-\frac{s_\Upsilon^2}{2n}\right]
 =\exp\left[-\frac{n[\Psi(n)-\Upsilon]^2}{32\sigma^2\Upsilon}\right].
\]
\item [(c)] Combine the bounds

The resulting conditional estimate is
\begin{equation}\label{eq:conditional}
\begin{aligned}
 \mathbb{P}(I_n^\Upsilon>\gamma_0)
 &\leq\exp\left[-\frac n2\{-u_\Upsilon-\ln(1-u_\Upsilon)\}\right]\\
 &\quad+\exp\left[-\frac{n[\Psi(n)-\Upsilon]^2}{32\sigma^2\Upsilon}\right].
 \end{aligned}
\end{equation}
For the first term of (\ref{eq:conditional}), we use (\ref{tempbound}) and
\begin{align*}
 \frac{nu_\Upsilon^2}{4}
 =\frac n4\left(\frac{\Psi(n)-\Upsilon}{2\mu\Psi(n)}\right)^2
 =\frac{n[\Psi(n)-\Upsilon]^2}{16\mu^2\Psi(n)^2}.
\end{align*}
to get 
\begin{align}
 \exp\left[-\frac n2\{-u_\Upsilon-\ln(1-u_\Upsilon)\}\right]  \nonumber \\
 \leq \exp\left[-\frac{n[\Psi(n)-\Upsilon]^2}{16\mu^2\Psi(n)^2}\right].
\end{align}
Finally, we arrive at 
\begin{align}\label{eachupsilonbound}
 \mathbb{P}(I_n^\Upsilon>\gamma_0)
 &\leq\exp\left[-\frac{n[\Psi(n)-\Upsilon]^2}{16\mu^2\Psi(n)^2}\right]\nonumber\\
 &\quad+\exp\left[-\frac{n[\Psi(n)-\Upsilon]^2}{32\sigma^2\Upsilon}\right].
 \end{align}
 \end{enumerate}
 
 \subsubsection{$\Upsilon$ in typical input-energy region $\mathcal F_n$}
 Restrict $\Upsilon$ in a typical region
 $\mathcal F_n \triangleq \{\Upsilon: \Upsilon\leq\frac{1+\mu}{2}\Psi(n)\}$.
then
\[
 \Psi(n)-\Upsilon\geq\frac{1-\mu}{2}\Psi(n),\, \  \,
 u_\Upsilon\geq u_*:=\frac{1-\mu}{4\mu}>0.
\]
The bound in (\ref{eachupsilonbound}) is
\begin{equation}\label{eq:typical}
\mathbb{P}(I_n^\Upsilon>\gamma_0)
 \leq e^{-an}+e^{-bn\Psi(n)}
\end{equation}
where $a=\frac{(1-\mu)^2}{64\mu^2},\, \,
 b=\frac{(1-\mu)^2}{128\sigma^2}.$
Both constants are independent of $n$ for fixed $\mu,\sigma$.
\subsubsection{Bounding the probability of $\mathcal F_n^c$}
Under $\mathcal N(\bm0,\mu \Psi(n) \bm{I}_n)$,
\[
 S_n=\frac{\|\bm X\|^2}{\mu \Psi(n)}\sim\chi_n^2.
\]
The event $\mathcal F_n^c$ corresponds to $S_n>nh$, where $h=(1+\mu)/(2\mu)>1$. For every $0<t<1/2$,
\[
 \mathbb{P}(S_n>nh)\leq\exp\left\{n\left[-th-\frac12\ln(1-2t)\right]\right\}.
\]
Let $f_h(t)=-th-\frac12\ln(1-2t)$. Then
\[
 f_h'(t)=-h+\frac{1}{1-2t},\qquad
 f_h''(t)=\frac{2}{(1-2t)^2}>0.
\]
Thus the unique minimum occurs at $t_*=(h-1)/(2h)\in(0,1/2)$. Substituting $1-2t_*=1/h$ gives
\[
 f_h(t_*)=-\frac{h-1}{2}+\frac{\ln h}{2}=-c_E,
 \qquad c_E=\frac{h-1-\ln h}{2}>0.
\]
The positivity follows from $\ln h<h-1$ for $h>1$. Therefore $ \mathbb{P}(S_n>nh)\leq e^{-nc_E}$.

As $\mathcal{A}_n$ has unconditioned probability $\Delta$. Under the truncated input law,
\[
  \mathbb{P}_{\overline P_n^{(n)}}(\mathcal F_n^c)
 =\frac{ \mathbb{P}(\mathcal F_n^c\cap\mathcal{A}_n)}{\Delta}
 \leq\frac{ \mathbb{P}(\mathcal F_n^c)}{\Delta}
 \leq\Delta^{-1}e^{-nc_E}.
\]
\subsubsection{Final bound}
 Let $p_n^{(0)}$ be the information-density tail under the joint truncated-input/channel law. By conditional expectation and~\eqref{eq:typical},
\begin{align}
 p_n^{(0)}
 &=\mathbb{E}_{\overline P_n^{(n)}}\left[\mathbb{P}(I_n^{\Upsilon}>\gamma_0\mid\Upsilon)\right]\\
 &\leq\mathbb{P}_{\overline P_n^{(n)}}(\mathcal F_n^c)
 +\mathbb{E}_{\overline P_n^{(n)}}\left[\bm1_{\mathcal F_n}\mathbb{P}(I_n^{\Upsilon}>\gamma_0\mid\Upsilon)\right]\\
 &\leq\Delta^{-1}e^{-c_En}+e^{-an}+e^{-bn\Psi(n)}.
\end{align}
with $a=\frac{(1-\mu)^2}{64\mu^2},\, \,
 b=\frac{(1-\mu)^2}{128\sigma^2}$.
\end{proof}

\subsection{Proof of Lemma 10}
\begin{proof}
\newcounter{TempEqCnt5}
\setcounter{TempEqCnt5}{\value{equation}}
\setcounter{equation}{119}
\begin{figure*}[!t]
\normalsize
\begin{align}\label{firsttermKL}
&\mathbb{E}_{W^n_2(\bm{Z}|\bm{X})\bar{P}_n^{(n)}(\bm{X})}\left(\log \left(1 + \frac{W^n_2(\bm{Z}|\bm{X})}{MKQ^{(n)}_{\bm{Z},1}(\bm{Z}) }\right)\right) \nonumber\\
= & \mathbb{E}_{W^n_2(\bm{Z}|\bm{x})\bar{P}_n^{(n)}(\bm{X})} \left[\log \left(1 + \frac{W^n_2(\bm{Z}|\bm{X})}{MKQ^{(n)}_{\bm{Z},1}(\bm{Z}) }\right)\bigg| \mathcal{E}\right] \mathbb{P}(\mathcal{E}) 
+   \mathbb{E}_{W^n_2(\bm{Z}|\bm{x})\bar{P}_n^{(n)}(\bm{X})} \left[\log \left(1 + \frac{W^n_2(\bm{Z}|\bm{X})}{MKQ^{(n)}_{\bm{Z},1}(\bm{Z}) }\right)\bigg| \mathcal{E}^c\right]\mathbb{P}(\mathcal{E}^c) \nonumber\\
\overset{(a)}{\leq} &\log \big( 1 + \frac{2^{\gamma_0}}{MK}\big) \cdot \mathbb{P}(\mathcal{E})  + \log \bigg( 1 + \frac{1}{MK}\cdot e^{\frac{n}{2\mu}}\big(1 + \frac{\mu \Psi(n)}{\sigma^2}\big)^{\frac{n}{2}}\bigg) \cdot \mathbb{P}(\mathcal{E}^c)\nonumber\\
\leq &\log \big( 1 + \frac{2^{\gamma_0}}{MK}\big) + \log \bigg( 1 + \frac{1}{MK}\cdot e^{\frac{n}{2\mu}}\big(1 + \frac{\mu \Psi(n)}{\sigma^2}\big)^{\frac{n}{2}}\bigg) \cdot \mathbb{P}(\mathcal{E}^c) \nonumber\\
\overset{(b)}{\leq} & \log \big( 1 + \frac{2^{\gamma_0}}{MK}\big) + dn p_n^{(0)}.
\end{align}
\hrulefill
\vspace*{4pt}
\end{figure*}
\setcounter{equation}{120}
The details can be found in (\ref{firsttermKL}), where step (a) follows from the definition of $\mathcal{E}$, and step (b) holds because of Lemma 9 and $\log \bigg( 1 + \frac{2}{MK}\cdot e^{\frac{n}{2\mu}}\big(1 + \frac{\mu \Psi(n)}{\sigma^2}\big)^{\frac{n}{2}}\bigg) \leq dn$ for some constant $d > 0$. 
\end{proof}

\subsection{Proof of Theorem 2}
\begin{proof}
We proceed with the equation (\ref{expection1}) of the main text. Since we have bounded the first term in Corollary 1, we just consider the second term in (\ref{expection1}). 
For that term,  as the distributions $Q^{(n)}_{\bm{Z},1}$ and $Q^{(n)}_{\bm{Z},0}$ are independent of $\bm{x}$, we have
\begin{align}\label{thirdKL}
\mathbb{E}_{\mathcal{C}}& \left[\int_{\bm{z}} Q^{(n)}_{\bm{Z},\mathcal{C}}(\bm{z}) \log \frac{Q^{(n)}_{\bm{Z},1}(\bm{z})}{Q^{(n)}_{\bm{Z},0}(\bm{z})}   d\bm{z} \right]\nonumber\\
= & \int_{\bm{z}} \left(\int_{\bm{x}} \frac{1}{MK} \sum_{s=1}^K\sum_{w=1}^M W^n_2(\bm{z}|\bm{x}_{sw}) d\bar{P}_n^{(n)}\right)\nonumber \\
\times & \log \frac{Q^{(n)}_{\bm{Z},1}(\bm{z})}{Q^{(n)}_{\bm{Z},0}(\bm{z})}  d\bm{z} \nonumber
\end{align}
\begin{align}
= & \int_{\bm{z}}\bar{Q}^{(n)}_{\bm{Z},1}(\bm{z}) \log \frac{Q^{(n)}_{\bm{Z},1}(\bm{z})}{Q^{(n)}_{\bm{Z},0}(\bm{z})}  d\bm{z}
\end{align}
From the definitions of the distributions at Willie,  
\begin{equation}
 Q_{\bm{Z},1}^{(n)}\sim \mathcal N(\bm0,[\sigma^2+\mu\Psi(n)]\mathbf \bm{I}_n),\,  \  \,
 Q_{\bm{Z},0}^{(n)}\sim \mathcal N(\bm0,\sigma^2\mathbf \bm{I}_n).
\end{equation}
 Taking their ratio gives
\begin{align}
 \log\frac{Q_{\bm{Z},1}^{(n)}(\bm z)}{Q_{\bm{Z},0}^{(n)}(\bm z)}
 &=-\frac n2\log\left(1+\frac{\mu\Psi(n)}{\sigma^2}\right)\nonumber\\
 &\quad+\frac{\mu\Psi(n)}{2\sigma^2[\sigma^2+\mu\Psi(n)]}
 \|\bm z\|^2\log e.\label{eq:llr}
\end{align}
The above two distributions is induced by $\bm{Z} = \bm{X} + \bm{N}$ with $\bm{X} \sim P_n^{(n)}$.
Let $\bar{E}_n=\mathbb{E}_{\bar P_n^{(n)}}[\|\bm X\|^2]$, we have
\begin{align}
  \bar{E}_n
  &= \frac{1}{\Delta} \int_{\mathcal{A}_n} \|\bm{x}\|^2 \, P_n^{(n)}(\mathrm{d}\bm{x}) \\
  &\leq \frac{1}{\Delta} \int_{\mathbb{R}^n} \|\bm{x}\|^2 \, P_n^{(n)}(\mathrm{d}\bm{x})
   = \frac{n \mu \Psi(n)}{\Delta}. \label{eq:energy}
\end{align}
From the following facts, 
\begin{align}
  \mathbb{E}_{\bar Q_{\bm{Z},1}^{(n)}}\!\left[\|\bm{Z}\|^2\right]
  &=  \mathbb{E}\!\left[\|\bm{X}\|^2\right] + 2 \mathbb{E}\!\left[\bm{X}^{\mathsf{T}}\bm{N}\right] + \mathbb{E}\!\left[\|\bm{N}\|^2\right] \nonumber\\
  &= \bar{E}_n + n\sigma^2, \label{eq:output-energy}
\end{align} 
\begin{align}
 \mathbb{D}(Q_{Z,1}^{(n)}\|Q_{Z,0}^{(n)})
 =\frac n2\left[\frac{\mu\Psi(n)}{\sigma^2}\log e
 -\log\left(1+\frac{\mu\Psi(n)}{\sigma^2}\right)\right],
\end{align}
\begin{equation}
 \bar E_n-n\mu\Psi(n)\leq n\mu\Psi(n)\frac{1-\Delta}{\Delta},
\end{equation}
we take expectation of (\ref{eq:llr}) with respect to $\bar{Q}^{(n)}_{\bm{Z},1}$ to get 
\begin{align}\label{upperbound2}
\int_{\bm{z}}&\bar{Q}^{(n)}_{\bm{Z},1}(\bm{z}) \log \frac{Q^{(n)}_{\bm{Z},1}(\bm{z})}{Q^{(n)}_{\bm{Z},0}(\bm{z})}  d\bm{z}\\
= &-\frac n2\log\left(1+\frac{\mu\Psi(n)}{\sigma^2}\right)
 +\frac{\mu\Psi(n)(n\sigma^2+\bar E_n)}
 {2\sigma^2[\sigma^2+\mu\Psi(n)]}\log e \nonumber \\
 = & \mathbb{D}(Q_{\bm{Z},1}^{(n)}\|Q_{\bm{Z},0}^{(n)})
 +\frac{\mu\Psi(n)[\bar E_n-n\mu\Psi(n)]}
 {2\sigma^2[\sigma^2+\mu\Psi(n)]}\log e \nonumber \\
\leq & \mathbb{D}(Q_{\bm{Z},1}^{(n)}\|Q_{\bm{Z},0}^{(n)})
 + r_n \nonumber 
\end{align}
with $ r_n$ in (\ref{r_ndefintion}) of the main text.
Finally, from Corollary 1 and (\ref{upperbound2}), we arrive at 
\begin{align}
\mathbb{E}_{\mathcal{C}}&  \left[\mathbb{D}(Q^{(n)}_{\bm{Z},\mathcal{C}}\| Q^{(n)}_{\bm{Z},0})\right] \nonumber \\
\leq & \log \big( 1 + \frac{2^{\gamma_0}}{MK}\big) + dn p_n^{(0)}  \nonumber \\
+ & \mathbb{D}(Q_{\bm{Z},1}^{(n)}\|Q_{\bm{Z},0}^{(n)})
 + r_n -\log \Delta . \nonumber
\end{align}
\end{proof}
\subsection{Proof of Lemma 11}\label{KLdifference0111}
\begin{proof}
Let $\mathcal C_1$ and $\mathcal C_2$ have $M$ codewords and differ only at index $k$. Define
\begin{align}
 Q_1(z)&= Q^{(n)}_{\bm{Z},\mathcal{C}_1} = \frac1M\sum_{i=1}^M W_2^n(z\mid x_i),\\
 Q_2(z)&= Q^{(n)}_{\bm{Z},\mathcal{C}_2} = \frac1M\left[W_2^n(z\mid x_k')+\sum_{i\ne k}W_2^n(z\mid x_i)\right],\\
 Q_0&= Q^{(n)}_{\bm{Z},0} = \mathcal N(0,\sigma^2\bm{I}_n).
\end{align}
Every codeword belongs to $\mathcal{A}_n$ and hence obeys
\[
 \|x_i\|^2\leq n\Psi(n),\qquad \|x_k'\|^2\leq n\Psi(n).
\]
Write $p_k=W_2^n(\cdot\mid x_k)$ and $p_k'=W_2^n(\cdot\mid x_k')$. Then $Q_1-Q_2=(p_k-p_k')/M$.

Adding and subtracting $\int Q_2\log Q_1$ gives
\begin{align}
 & \mathbb{D}(Q_1\|Q_0)-\mathbb{D}(Q_2\|Q_0)\nonumber\\
 &=\int(Q_1-Q_2)\log Q_1\,dz  \nonumber \\
 - & \int(Q_1-Q_2)\log Q_0\,dz \nonumber \\ 
 - & \mathbb{D}(Q_2\|Q_1).
\end{align}
Consequently,
\begin{equation}\label{eq:decomposition}
 |\mathbb{D}(Q_1\|Q_0)-\mathbb{D}(Q_2\|Q_0)|\leq T_1+T_2+T_3,
\end{equation}
where
\begin{align}
 T_1&=\left|\int(Q_1-Q_2)\log Q_1\,dz\right|,\label{T_1expression}\\
 T_2&=\left|\int(Q_1-Q_2)\log Q_0\,dz\right|,\label{T_2expression}\\
 T_3&=\mathbb{D}(Q_2\|Q_1).\label{T_3expression}
\end{align}
We will provide bounds for the three terms in (\ref{eq:decomposition}) separately.  Before that, we provide a proposition for preparation.
\subsection{A proposition}\label{techniquelemma1}
\begin{Proposition}
 For any $\bm{z}\in\mathbb{R}^n$,
\begin{align}
\big|\log  Q_1(z)\big|
\;\leq & \;
\max_{i\in[M]}\frac{\|\bm{z}-\bm{x}_i\|^2}{2\sigma^2}\cdot\log e \nonumber \\
\;+ & \;\Big|\frac{n}{2}\log(2\pi\sigma^2)\Big|.
\end{align}
\end{Proposition}
\begin{proof}
Rewrite the quantity inside the absolute value as
\begin{align}\label{logpdf10}
\log  Q_1(z)
= &\log\Big(\frac{1}{M}\sum_{i=1}^{M}
\exp\Big(-\frac{\|\bm{z}-\bm{x}_i\|^2}{2\sigma^2}\Big)\Big) \nonumber \\
- &\frac{n}{2}\log(2\pi\sigma^2).
\end{align}

By the triangle inequality,
\begin{align}
\big|\log  Q_1(z)\big|
\leq &
\Big|\log\Big(\frac{1}{M}\sum_{i=1}^{M}
\exp\Big(-\frac{\|\bm{z}-\bm{x}_i\|^2}{2\sigma^2}\Big)\Big)\Big|\nonumber \\
+ &\Big|\frac{n}{2}\log(2\pi\sigma^2)\Big|.
\end{align}
It remains to bound the first term. Since $\sigma^2>0$, each exponent satisfies
$-\|\bm{z}-\bm{x}_i\|^2/(2\sigma^2)\leq 0$, hence every summand lies in $(0,1]$,
and therefore
\begin{align}
0<\frac{1}{M}\sum_{i=1}^{M}\exp\Big(-\frac{\|\bm{z}-\bm{x}_i\|^2}{2\sigma^2}\Big)\leq 1,
\end{align}
which implies that the logarithm is non-positive. Moreover, by monotonicity of the
logarithm and the fact that the average dominates the minimum,
\begin{align}
\log\Big(\frac{1}{M}\sum_{i=1}^{M}&
\exp\Big(-\frac{\|\bm{z}-\bm{x}_i\|^2}{2\sigma^2}\Big)\Big) \nonumber \\
\;\geq\;
&\log\min_{i\in[M]}\exp\Big(-\frac{\|\bm{z}-\bm{x}_i\|^2}{2\sigma^2}\Big) \nonumber \\
= &-\max_{i\in[M]}\frac{\|\bm{z}-\bm{x}_i\|^2}{2\sigma^2} \log e,
\end{align}
where the last equality uses $\log\min_i f_i = \min_i \log f_i$. Combining the two displays gives
\begin{align}
\Big|\log\Big(\frac{1}{M}\sum_{i=1}^{M}&
\exp\Big(-\frac{\|\bm{z}-\bm{x}_i\|^2}{2\sigma^2}\Big)\Big)\Big|\nonumber \\
&\;\leq\;
\max_{i\in[M]}\frac{\|\bm{z}-\bm{x}_i\|^2}{2\sigma^2}\cdot\log e,
\end{align}
\end{proof}
\subsubsection{Bounding the first term}
From (\ref{T_1expression}) and $Q_1-Q_2=(p_k-p_k')/M$,
\begin{equation}\label{eq:T1-start}
 T_1\leq\frac1M\left[\mathbb{E}_{Z\mid x_k}|\log Q_1(Z)|+
 \mathbb{E}_{Z\mid x_k'}|\log Q_1(Z)|\right].
\end{equation}
Under $Z=x_k+N$, where $N\sim\mathcal N(0,\sigma^2\bm{I}_n)$, bound pointwise:
\begin{align}
 \max_i\|Z-x_i\|^2
 &=\max_i\|N+x_k-x_i\|^2\notag\\
 &\leq2\|N\|^2+2\max_i\|x_k-x_i\|^2\notag\\
 &\leq2\|N\|^2+8n\Psi(n).\label{eq:pointwise}
\end{align}
The first inequality uses $\|a+b\|^2\leq2\|a\|^2+2\|b\|^2$. The second uses
\[
 \|x_k-x_i\|\leq\|x_k\|+\|x_i\|\leq2\sqrt{n\Psi(n)}.
\]
Taking expectations after this bound is valid and yields
\begin{equation}
 \mathbb{E}_{Z\mid x_k}\max_i\|Z-x_i\|^2\leq2n\sigma^2+8n\Psi(n).
\end{equation}
The same estimate holds under $Z=x_k'+N$, since $x_k'$ satisfies the same power constraint. Independence between the distances indexed by $i$ is not required.

Combining these estimates with \eqref{eq:T1-start} gives
\begin{equation}\label{eq:T1}
 T_1\leq\frac nM\left[
 \frac{2\sigma^2+8\Psi(n)}{\sigma^2}\log e
 +|\log(2\pi\sigma^2)|\right].
\end{equation}
\subsubsection{Bounding the second term}
Since
\[
 |\log Q_0(z)|\leq\frac{\|z\|^2}{2\sigma^2}\log e
 +\frac n2|\log(2\pi\sigma^2)|,
\]
and
\[
 \mathbb{E}_{Z\mid x_k}\|Z\|^2=n\sigma^2+\|x_k\|^2
 \leq n\sigma^2+n\Psi(n),
\]
with the same bound under $x_k'$, the single-codeword difference gives
\begin{equation}\label{eq:T2}
T_2\leq\frac nM\left[
 \frac{\sigma^2+\Psi(n)}{\sigma^2}\log e
 +|\log(2\pi\sigma^2)|\right].
\end{equation}
\subsubsection{Bounding the third term}
Write the common mixture of all unchanged codewords as $R$. For $M>1$,
\[
 Q_1=\frac1M p_k+\frac{M-1}{M}R,\qquad
 Q_2=\frac1M p_k'+\frac{M-1}{M}R.
\]
The convexity of KL divergence yields
\[
 T_3\leq\frac1M \mathbb{D}(p_k'\|p_k).
\]
As we have
\[
 \mathbb{D}(\mathcal N(x_k',\sigma^2\bm{I}_n)\|\mathcal N(x_k,\sigma^2\bm{I}_n))
 =\frac{\|x_k'-x_k\|^2}{2\sigma^2}\log e.
\]
Because $\|x_k'-x_k\|^2\leq 2(\|x_k'\|^2 + \|x_k\|^2)\leq4n\Psi(n)$,
\begin{equation}\label{eq:T3}
T_3\leq\frac{2n\Psi(n)}{M\sigma^2}\log e.
\end{equation}
\subsubsection{Bounding $T_1 +T_2+ T_3$}
Adding the bounds above leads to 
\begin{equation}\label{eq:theta}
T_1 +T_2+ T_3 \leq \frac{n}{M}\Theta(n)
\end{equation}
where
\begin{equation}
\Theta(n)= \frac{3\sigma^2+11\Psi(n)}{\sigma^2}\log e
 +2|\log(2\pi\sigma^2)|.
\end{equation}
\end{proof}
\subsection{Proof of Lemma 13}
\begin{proof}
For each key \(s\), define
\[
A_s:=\frac1{M_{\rm raw}}
\sum_{w=1}^{M_{\rm raw}}\varepsilon_{sw}^{(2)}.
\]
The Type-II ensemble bound in Lemma 6, applied with
\(M_n=M_{\rm raw}\), then gives
\[
\mathbb E_{\mathcal C}\left[
\frac1{M_{\rm raw}}
\sum_{w=1}^{M_{\rm raw}}\varepsilon_{sw}^{(2)}
\right]\leq n^{-c_0}R_{Y,n}.
\]
By Markov's inequality,
\[
\mathbb P\{A_s>
\lambda_1n^{-c_0}R_{Y,n}\}
\leq\lambda_1^{-1}.
\]
Each \(A_s\) depends only on its independently
generated key subcode. Therefore,
\[
\mathbb P(\mathcal E_1)
=\prod_{s=1}^{K}
\mathbb P\{A_s\leq\lambda_1n^{-c_0}R_{Y,n}\}
\geq(1-\lambda_1^{-1})^K.
\]

For each fixed subset \(\mathcal I\)
of size \(KM_{\rm fin}\), put
\[
B_{\mathcal I}:=
\mathbb D(Q_{\bm Z,\bm C}^{\mathcal I}
\|Q_{\bm Z,0}^{(n)}).
\]
Its selected codewords remain independent,
so Theorem 2 gives
\(\mathbb E_{\mathcal C}B_{\mathcal I}\leq b_n\),
where
\[
b_n:=G_n+
\log\left(1+\frac{2^{\gamma_0}}{KM_{\rm fin}}\right)
+dn p_n^{(0)}+r_n-\log\Delta.
\]
Since \(\mathcal E_2\) requires
\(B_{\mathcal I}\leq b_n+\lambda_3\)
for every such subset,
\[
\mathcal E_2^c
\subseteq
\bigcup_{|\mathcal I|=KM_{\rm fin}}
\{B_{\mathcal I}>
\mathbb E_{\mathcal C}B_{\mathcal I}+\lambda_3\}.
\]
There are $\binom{KM_{\rm raw}}{KM_{\rm fin}}$
possible subsets of size \(KM_{\rm fin}\).
For each fixed subset, Lemma 12 gives
\begin{align}
\mathbb P & \left\{
\mathbb D 
\left(Q_{\bm Z,\bm C}^{\mathcal I}
\middle\|Q_{\bm Z,0}^{(n)}\right)
>
\mathbb E_{\mathcal C}
\mathbb D 
\left(Q_{\bm Z,\bm C}^{\mathcal I}
\middle\|Q_{\bm Z,0}^{(n)}\right)
+\lambda_3
\right\} \nonumber\\
\leq &
\exp\left(
-\frac{2KM_{\rm fin}\lambda_3^2}{n^2\Theta_n^2}
\right).
\end{align}
Since \(\mathbb H_b\) uses base-2 logarithms,
\[
\binom{KM_{\rm raw}}{KM_{\rm fin}}
\leq
2^{KM_{\rm raw}\mathbb H_b(\widehat\lambda_{2,n})}.
\]
Define
\[
q_n:=
\frac{2\widehat\lambda_{2,n}\lambda_3^2}
     {n^2\Theta_n^2}
-(\ln 2)\mathbb H_b(\widehat\lambda_{2,n}).
\]
The union bound therefore yields
\[
\mathbb P(\mathcal E_2^c)
\leq \exp(-KM_{\rm raw}q_n).
\]
Combining this estimate with
\[
\mathbb P(\mathcal E_1)
\geq(1-\lambda_1^{-1})^K,
\]
we obtain
\begin{align}
\mathbb P(\mathcal E_1\cap\mathcal E_2)
&\geq
\mathbb P(\mathcal E_1)
+\mathbb P(\mathcal E_2)-1\nonumber\\
&\geq
(1-\lambda_1^{-1})^K
-\exp(-KM_{\rm raw}q_n).
\end{align}
In particular, the intersection has strictly
positive probability whenever
\begin{equation}\label{capevents2}
\lambda_1^{-1}
+\exp(-M_{\rm raw}q_n)<1.
\end{equation}
Indeed, this condition implies
\[
1-\lambda_1^{-1}
>\exp(-M_{\rm raw}q_n).
\]
\end{proof}

\subsection{Proof of Theorem 3}
The proof consists of five steps (1)-(5). First, we establish the maximal-error reliability of the expurgated code. Second, we verify the covertness constraint by showing that the residual terms are dominated by the available covertness margin. Third, we derive the first- and second-order achievable throughput asymptotics and show that expurgation does not affect these coefficients. Fourth, we establish the strict codebook-existence condition, explicitly accounting for integer rounding and the binary-entropy penalty. Finally, we derive sufficient secret-key sizes from the resolvability condition and characterize their asymptotic behavior in the different regimes determined by \(\bar A_\mu\).
\begin{proof} 
(1) We first consider the reliability. 
Choose a realization in the event
\(\mathcal E_1\cap\mathcal E_2\).
For each key \(s\), order the original Type-II
error probabilities as
\[
a_{s,1}\leq a_{s,2}\leq\cdots\leq
a_{s,M_{\rm raw}}.
\]
Retain the \(M_{\rm fin}\) messages with the
smallest Type-II error probabilities, and
relabel them as \(1,\ldots,M_{\rm fin}\).
On \(\mathcal E_1\),
\[
\sum_{j=1}^{M_{\rm raw}}a_{s,j}
\leq M_{\rm raw}\lambda_1n^{-c_0}R_{Y,n}.
\]
Every discarded message has Type-II error
probability at least \(a_{s,M_{\rm fin}}\). Therefore,
\[
(M_{\rm raw}-M_{\rm fin})a_{s,M_{\rm fin}}
\leq\sum_{j=1}^{M_{\rm raw}}a_{s,j}
\leq M_{\rm raw}\lambda_1n^{-c_0}R_{Y,n}.
\]
Consequently, the maximum Type-II
error probability among retained messages is
at most
\[
a_{s,M_{\rm fin}}
\leq
\frac{\lambda_1}
     {1-\widehat\lambda_{2,n}}
n^{-c_0}R_{Y,n}.
\]

Use the threshold decoder on the retained
messages, treating every discarded-message
output as an error. For each retained message,
the probability of decoding error does not
increase. Thus the expurgated code satisfies
\[
\max_{s,w}\varepsilon_{sw}
\leq
\epsilon_1+
\frac{\lambda_1}
     {1-\widehat\lambda_{2,n}}
n^{-c_0}R_{Y,n}.
\]

Let \(\mathcal I_{\rm fin}\) be the set of
retained message-key indices. Since exactly
\(M_{\rm fin}\) messages are retained for every key,
\[
|\mathcal I_{\rm fin}|=KM_{\rm fin}.
\]
Under uniform messages and a uniform key, the
final output distribution is
\[
Q_{\bm Z,\mathcal C_{\rm fin}}^{(n)}
=\frac1{KM_{\rm fin}}
\sum_{(s,w)\in\mathcal I_{\rm fin}}
W_2^n(\cdot\mid\bm x_{sw})
=Q_{\bm Z,\bm C}^{\mathcal I_{\rm fin}}.
\]
The event \(\mathcal E_2\) therefore supplies the
required covertness bound for the expurgated code.
For \(\lambda_2=1-n^{-11}\),
\[
\widehat\lambda_{2,n}\leq\lambda_2,
\qquad
1-\widehat\lambda_{2,n}\geq n^{-11}.
\]
With \(c_0=13\), \(\lambda_1=n\), and
\(R_{Y,n}\leq\Delta^{-1}\), we obtain
\[
\frac{\lambda_1}
     {1-\widehat\lambda_{2,n}}
n^{-c_0}R_{Y,n}
\leq\frac{1}{\Delta n}.
\]

Fix \(\epsilon\in(0,1/2)\) and choose
\[
\epsilon_1(n):=\epsilon-\frac{2}{n}
\]
for sufficiently large \(n\).
Since \(\Delta\to1\), eventually \(\Delta\geq1/2\).
Therefore,
\[
\max_{s,w}\varepsilon_{sw}
\leq\epsilon_1(n)+\frac{1}{\Delta n}
\leq\epsilon.
\]
Together with Lemma 4, this proves that the
maximal decoding error probability of the
expurgated code is at most \(\epsilon\).

(2) The sufficient covertness inequality (74) is
\begin{equation}\label{eq:full}
G_n+\log\left(1+\frac{2^{\gamma_0}}{KM_{\rm fin}}\right)
 +dn p_n^{(0)}+r_n-\log\Delta+\lambda_3\leq\delta.
\end{equation}
Set
$S_n=\delta-G_n $ and 
\begin{align}
R_n=\log\left(1+\frac{2^{\gamma_0}}{KM_{\rm fin}}\right)
+dn p_n^{(0)}+t_n+\lambda_3
\end{align}
where $t_n = r_n-\log\Delta$.
The final verification of covertness is $R_n\le S_n$ for every sufficiently large $n$. 
We analyze each term of $R_n$, separately. 
The raw key-size condition gives
\[
\frac{2^{\gamma_0}}{M_{\rm raw}K}
\leq n^{-c_5}.
\]
Since
\(M_{\rm fin}=\widehat\lambda_{2,n}M_{\rm raw}\)
and \(\widehat\lambda_{2,n}\geq1/2\) eventually,
\begin{equation}\label{asymptoticquantities}
\begin{aligned}
\log\left(1+\frac{2^{\gamma_0}}{KM_{\rm fin}}\right)
&\leq
\frac{\log e}{\widehat\lambda_{2,n}n^{c_5}}\\
&\leq2(\log e)n^{-c_5}.
\end{aligned}
\end{equation}
Here we used \(\log(1+u)\leq u\log e\).
With \(c_5=3\), this contribution is
\(O(n^{-3})=o(n^{-1/2})\).

From Lemma 9,
\[
 p_n^{(0)}\leq\Delta^{-1}e^{-c_En}+e^{-an}+e^{-bn\Psi(n)},
 \]
 For $\Psi(n)=c_4n^{-1/2}$ and fixed $\mu$, the last exponent is $bc_4\sqrt n$. Since $\Delta\to1$, there exist constants $A,c>0$ independent of $n$ such that $p_n^{(0)}\leq Ae^{-c\sqrt n}$ eventually. In particular,
\begin{align}\label{boundbound}
 \frac{dn p_n^{(0)}}{n^{-1/2}}
 \leq dA n^{3/2}e^{-c\sqrt n}\longrightarrow0
\end{align}
implies  $dn p_n^{(0)}=o(n^{-1/2})$.
The proof of $r_n-\log\Delta\le4(L+\log e)e^{-nc_-(\mu)}$ can be found in Section \ref{boundonr_n-logdelta}, and since $L$ and $c_-(\mu)>0$ are independent of $n$ for fixed $\mu$, it is obvious that 
\begin{align}\label{boundbound2}
t_n = r_n-\log\Delta =o(n^{-1/2})
\end{align} 
for sufficiently large $n$.

Finally, with $\lambda_3=n^{-c_3}$ and $c_3=4$,
$\lambda_3=o(n^{-1/2})$.

Combining the above analysis gives
\[
R_n=o(n^{-1/2}).
\]
For $G_n$, Taylor's theorem with power level $
\Psi(n)=\frac{2\sigma^2}{\mu}
\sqrt{\frac{\delta\ln2}{n}}$ 
yields
\begin{align}
G_n=\delta-\frac43\delta^{3/2}\sqrt{\ln2}\,n^{-1/2}
+O(n^{-1}),
\end{align}
which further leads to 
\begin{align}\label{S_nbound}
S_n=C_\delta n^{-1/2}+O(n^{-1}),\, \  \,
C_\delta=\frac43\delta^{3/2}\sqrt{\ln2}>0.
\end{align}
Therefore, 
\begin{align}
R_n\le \frac{C_\delta}{4}n^{-1/2}\le S_n
\quad\text{for all sufficiently large }n.
\end{align}
Thus \eqref{eq:full} follows.

(3) From the definitions of $\mathsf{C}(\mu,\Psi(n))$, $\bar{\mathsf{C}}(\mu^2 \Psi(n),\mu\Psi(n))$ and $\mathsf V(\Upsilon,\mu\Psi(n))$  of the main text, using $\Psi(n) = c_4\cdot \frac{\sqrt{n}}{n}$, we get 
\begin{align}
n\mathsf{C}(\mu, \Psi(n)) = & \frac{n}{2}\log (1 + \mu \Psi(n)) \nonumber\\
 = & \frac{n}{2} \log (1 + \mu c_4\cdot \frac{\sqrt{n}}{n})\nonumber\\
 = & \frac{\mu c_4 \sqrt{n}}{2} \log e + O(1),
\end{align}
\begin{align}
n\bar{\mathsf{C}}(\mu^2 \Psi(n),\mu\Psi(n)) 
= & n\cdot\frac{\mu^2 \Psi(n) - \mu \Psi(n)}{2(1+\mu \Psi(n))} \log e \nonumber \\
 = & \frac{1}{2}(\mu^2 -\mu) c_4  \sqrt{n}\log e + o(\sqrt{n}),
\end{align}
\begin{align}
\mathsf V(\Psi(n),\mu\Psi(n)) 
= &\frac{2\Psi(n)+\mu^2\Psi(n)^2}{2(1+\mu\Psi(n))^2}(\log e)^2 \nonumber\\
= & \frac{1}{\sqrt{n}} c_4 \log^2 e + o(\frac{1}{\sqrt{n}}).
\end{align}
By the uniform estimate in the proof of Lemma 4,
$\beta_n=O(n^{-1/2})$. Therefore
\[
 z_n:=\epsilon_1-\beta_n=\epsilon+O(n^{-1/2}),
\]
and $0<z_n<1/2$ for all sufficiently large $n$. The mean-value theorem applied to $\Phi^{-1}$ near the fixed point $\epsilon$ gives
\[
 \Phi^{-1}(z_n)=\Phi^{-1}(\epsilon)+O(n^{-1/2}).
\]
Write $v_n=\sqrt{n\mathsf V(\Psi(n),\mu\Psi(n))}$. Since $v_n=\Theta(n^{1/4})$, it follows that
\begin{equation}
 v_n\Phi^{-1}(z_n)
 =v_n\Phi^{-1}(\epsilon)+O(n^{-1/4}).
\end{equation}
Thus, we have 
\begin{align}\label{logMsecondfinal}
 & \log M_{\mathrm{raw}} \\
 &= \tfrac{\mu^2 c_4}{2}\sqrt{n}\lg e
    + \sqrt{c_4}n^{1/4}\!\lg e\,\Phi^{-1}(z_n)
    + o(n^{1/4}) \nonumber \\[-1pt]
 &= \tfrac{\mu^2 c_4}{2}\sqrt{n}\lg e
    + \sqrt{c_4}n^{1/4}\!\lg e\,\Phi^{-1}(\epsilon)
    + o(n^{1/4}).\nonumber
\end{align}
Since
\[
M_{\rm fin}
=\widehat\lambda_{2,n}M_{\rm raw},
\]
we obtain
\[
\log M_{\rm fin}
=\log M_{\rm raw}
+\log\widehat\lambda_{2,n}
=\log M_{\rm raw}+o(1).
\]
Thus the same first- and second-order
throughput coefficients hold for the final code.
 
(4) We next verify the strict codebook-existence
condition
\[
\lambda_1^{-1}+\exp(-M_{\rm raw}q_n)<1,
\]
where
\[
q_n:=
\frac{2\widehat\lambda_{2,n}\lambda_3^2}
     {n^2\Theta_n^2}
-(\ln 2)\mathbb H_b(\widehat\lambda_{2,n}).
\]
Here \(\mathbb H_b\) denotes binary entropy
with base-2 logarithms.

For our chosen parameters,
\[
\lambda_1=n,\qquad
\lambda_2=1-n^{-11},\qquad
\lambda_3=n^{-4},
\]
and
\[
M_{\rm fin}=\lfloor\lambda_2M_{\rm raw}\rfloor,
\qquad
\widehat\lambda_{2,n}=\frac{M_{\rm fin}}{M_{\rm raw}}.
\]

First, we quantify the rounding correction.
By the definition of the floor function,
there exists \(\eta_n\in[0,1)\) such that
\[
M_{\rm fin}=(1-n^{-11})M_{\rm raw}-\eta_n.
\]
Consequently,
\[
u_n:=1-\widehat\lambda_{2,n}
=n^{-11}+\frac{\eta_n}{M_{\rm raw}}.
\]

The message-size expansion established in this
proof gives
\[
\log M_{\rm raw}
=a\sqrt n+O(n^{1/4}),
\qquad
a:=\mu\sigma^2\sqrt{\delta\log e}>0.
\]
Hence
\[
M_{\rm raw}^{-1}=o(n^{-11}).
\]
In particular, for all sufficiently large \(n\),
\[
n^{-11}\leq u_n\leq2n^{-11},
\qquad
\widehat\lambda_{2,n}\geq\frac12.
\]

Next, by the symmetry of binary entropy,
\[
\mathbb H_b(\widehat\lambda_{2,n})
=\mathbb H_b(u_n).
\]
For \(0<u<1\),
\[
\mathbb H_b(u)
\leq u\log\frac{e}{u}.
\]
The function \(u\log(e/u)\) is increasing on
\((0,1)\). Therefore, eventually,
\begin{align*}
\mathbb H_b(\widehat\lambda_{2,n})
&\leq
2n^{-11}\log\frac{e}{2n^{-11}}\\
&\leq
2n^{-11}\bigl(11\log n+\log e\bigr).
\end{align*}
Thus,
\[
(\ln 2)\mathbb H_b(\widehat\lambda_{2,n})
=O(n^{-11}\log n)=o(n^{-10}).
\]

Since \(\Psi(n)=c_4n^{-1/2}\) is bounded,
the expression defining \(\Theta_n\) gives
a constant \(\Theta_{\max}>0\), independent
of \(n\), such that
\[
\Theta_n\leq\Theta_{\max}.
\]
Put $\nu:=\Theta_{\max}^{-2}>0 $.
Using \(\widehat\lambda_{2,n}\geq1/2\) and
\(\lambda_3=n^{-4}\), we obtain
\[
\frac{2\widehat\lambda_{2,n}\lambda_3^2}
     {n^2\Theta_n^2}
\geq \nu n^{-10}.
\]
The entropy term is \(o(n^{-10})\), so,
for all sufficiently large \(n\),
\begin{equation}\label{thridtermquantity}
q_n\geq\frac{\nu}{2}n^{-10}>0.
\end{equation}

Finally, the positive leading message-size
coefficient implies
\[
M_{\rm raw}\geq2^{a\sqrt n/2}
\]
for all sufficiently large \(n\). Hence
\[
M_{\rm raw}q_n
\geq
\frac{\nu}{2}n^{-10}2^{a\sqrt n/2}
\longrightarrow\infty.
\]
It follows that
\[
\lambda_1^{-1}+\exp(-M_{\rm raw}q_n)
\leq
\frac1n+
\exp\left(
-\frac{\nu}{2}M_{\rm raw}n^{-10}
\right)
\longrightarrow0.
\]
Therefore,
\[
\lambda_1^{-1}+\exp(-M_{\rm raw}q_n)<1
\]
for all sufficiently large \(n\), proving
the strict codebook-existence condition.

(5) From the condition (c) of the main text, 
\[
 x_n=\gamma_0+c_5\log n-\log M_{\rm raw},\qquad
 K_n=\left\lceil 2^{x_n}\right\rceil.
\]
will satisfy (76). If $x_n\leq0$, $K_n=1$. If $x_n>0$, then $2^{x_n}\leq K_n<2^{x_n}+1<2^{x_n+1}$. With $[x]^+=\max\{0,x\}$, this proves
\[
 [x_n]^+\leq\log K_n<[x_n]^++1.
\]
Thus, we have
\begin{equation}\label{eq:keyexact}
\log K_n=
\left[
n\widetilde{\mathsf C}_n
-n\widehat{\mathsf C}_n
-\log M_{\rm raw}
+c_5\log n
\right]^++O(1).
\end{equation}
where the rounding correction is less than one bit. Define
\[
v_n:=\sqrt{n\mathsf V(\Psi(n),\mu\Psi(n))},
\qquad
z_n:=\epsilon_1(n)-\beta_n.
\]
Write $\mathsf{C}_n,\overline{\mathsf{C}}_n,\widetilde{\mathsf{C}}_n,\widehat{\mathsf{C}}_n$
for the corresponding per-use quantities. Define the
message-size remainder by
\[
\ell_n:=
\log M_{\rm raw}-n\mathsf{C}_n-n\overline {\mathsf{C}}_n-v_n\Phi^{-1}(z_n).
\]
Thus,
\[
\log M_{\rm raw}
=n\mathsf{C}_n+n\overline{\mathsf{C}}_n+v_n\Phi^{-1}(z_n)+\ell_n.
\]
For the explicit threshold-based message-size choice,
including integer rounding, \(\ell_n=O(\log n)\).
Now set
\begin{align*}
F_n&:=n(\widetilde{\mathsf{C}}_n-\widehat{\mathsf{C}}_n-\mathsf{C}_n-\overline {\mathsf{C}}_n),\\
J_n&:=-v_n\Phi^{-1}(z_n),\\
L_n&:=c_5\log n-\ell_n.
\end{align*}
These definitions give the exact identity
\[
F_n+J_n+L_n
=n\widetilde{\mathsf{C}}_n-n\widehat{\mathsf{C}}_n-\log M_{\rm raw}+c_5\log n
=x_n.
\]
Consequently,
\begin{equation}\label{eq:keyfull}
\log K_n=[F_n+J_n+L_n]^++O(1).
\end{equation}
Using $\log(1+t)=t\log e+O(t^2)$ as $t\to0$, and $n\Psi(n)^2=O(1)$, gives
\begin{align*}
 n(\widetilde{\mathsf{C}}_n-\mathsf{C}_n)
 &=\frac{\mu c_4}{2}(\sigma^{-2}-1)\sqrt n\log e+O(1),\\
 -n\widehat{\mathsf{C}}_n
 &=\frac{(1-\mu)c_4}{2\sigma^2}\sqrt n\log e+O(1),\\
 -n\overline {\mathsf{C}}_n
 &=\frac{(\mu-\mu^2)c_4}{2}\sqrt n\log e+O(1).
\end{align*}
Adding the coefficients,
\[
 \mu(\sigma^{-2}-1)+(1-\mu)\sigma^{-2}+(\mu-\mu^2)
 =\sigma^{-2}-\mu^2.
\]
Consequently,
\begin{align*}
 F_n
 &=\frac{1-\mu^2\sigma^2}{2\sigma^2}c_4\sqrt n\log e+O(1)\\
 &=\frac{1-\mu^2\sigma^2}{\mu}\sqrt{\delta n\log e}+O(1).
\end{align*}
Define
\begin{align}
\bar A_\mu
&:=\frac{1-\mu^2\sigma^2}{\mu}
   \sqrt{\delta\log e},\\
B
&:=\sigma\sqrt{\frac{2}{\mu}}\,
   \delta^{1/4}(\log e)^{3/4}
   \Phi^{-1}(1-\epsilon).
\end{align}
Since \(\epsilon\in(0,1/2)\), we have \(B>0\).
The preceding expansion gives
\[
F_n=\bar A_\mu\sqrt n+O(1).
\]
Moreover, the uniform Berry--Esseen estimate and
the choice \(\epsilon_1(n)=\epsilon-2/n\) imply
\[
z_n=\epsilon_1(n)-\beta_n
=\epsilon+O(n^{-1/2}).
\]
Because \(\epsilon\) is fixed in \((0,1/2)\),
\[
\Phi^{-1}(z_n)
=\Phi^{-1}(\epsilon)+O(n^{-1/2}).
\]
Using the dispersion expansion,
\[
v_n=\sqrt{c_4}\,n^{1/4}\log e+o(n^{1/4}),
\]
we obtain
\begin{align}
J_n
&=-v_n\Phi^{-1}(z_n)\nonumber\\
&=\sqrt{c_4}\,\Phi^{-1}(1-\epsilon)
  n^{1/4}\log e+o(n^{1/4})\nonumber\\
&=Bn^{1/4}+o(n^{1/4}),
\end{align}
where
\[
c_4=\frac{2\sigma^2}{\mu}\sqrt{\delta\ln 2}.
\]
For the explicit threshold-based message-size
construction, \(\ell_n=O(\log n)\). Hence
\[
L_n=c_5\log n-\ell_n
=O(\log n)=o(n^{1/4}).
\]
Combining these estimates with
\(x_n=F_n+J_n+L_n\), we obtain
\[
x_n=\bar A_\mu\sqrt n
    +Bn^{1/4}+o(n^{1/4}).
\]
Therefore,
\begin{equation}\label{keycasesfinal}
\begin{aligned}
\log K_n
=\Big[
&\bar A_\mu\sqrt n
 +Bn^{1/4}\\
&+o(n^{1/4})
\Big]^++O(1),
\end{aligned}
\end{equation}
where the \(O(1)\) term accounts for integer
rounding and is less than one bit.
For fixed \(\mu\in(1/2,1)\) and \(\epsilon\in(0,1/2)\),
the preceding estimates yield
\[
x_n=\bar A_\mu\sqrt n+Bn^{1/4}+o(n^{1/4}).
\]
Together with the integer rounding bound, this implies
\[
\log K_n=
\begin{cases}
\bar A_\mu\sqrt n+Bn^{1/4}+o(n^{1/4}),
&\bar A_\mu>0,\\[3pt]
Bn^{1/4}+o(n^{1/4}),
&\bar A_\mu=0,\\[3pt]
0\quad\text{for all sufficiently large }n,
&\bar A_\mu<0.
\end{cases}
\]
In the last case, the negative \(\sqrt n\) term dominates
the smaller-order terms, so \(x_n<0\) eventually and
\(K_n=1\) exactly. These key sizes are sufficient for
the proposed construction.
\end{proof}

\subsection{Proof of Lemma 14}
\begin{proof}
Define $(S,W)$ as uniform over the index pairs and then set $\bm X=\bm x_{SW}$.
Then
\[
\mathbb E\|\bm X\|^2
=n\overline P_{\bm{c}}.
\]
Write
\[
\bm Z=\bm X+\bm N,
\qquad
\bm N\sim\mathcal N(\bm0,\sigma^2\bm I_n),
\]
where \(\bm N\) is independent of \(\bm X\).
Therefore,
\[
\mathbb E\|\bm Z\|^2
=n(\sigma^2+\overline P_{\bm{c}}).
\]
Define
\[
Q_G:=
\mathcal N\left(
\bm0,
(\sigma^2+\overline P_{\bm{c}})\bm I_n
\right).
\]
Nonnegativity of
\(\mathbb D(Q_{\bm Z,\bm c}^{(n)}\|Q_G)\),
together with the second-moment identity,
gives
\[
h(\bm Z)
\leq
\frac n2
\log\left(
2\pi e(\sigma^2+\overline P_{\bm c})
\right).
\]
This argument does not require
\(\mathbb E\bm X=\bm0\).

Since
\(Q_{\bm Z,0}^{(n)}
=\mathcal N(\bm0,\sigma^2\bm I_n)\),
we have
\begin{align*}
&\mathbb D
\left(Q_{\bm Z,\bm c}^{(n)}
\middle\|Q_{\bm Z,0}^{(n)}\right)\\
&\quad=
-h(\bm Z)
+\frac n2\log(2\pi\sigma^2)
+\frac{\mathbb E\|\bm Z\|^2}
       {2\sigma^2}\log e\\
&\quad\geq
\frac n2
\left[
\frac{\overline P_{\bm c}}{\sigma^2}
-\ln\left(
1+\frac{\overline P_{\bm c}}{\sigma^2}
\right)
\right]\log e.
\end{align*}

Define
\[
f_n(p):=
\frac n2
\left[
\frac p{\sigma^2}
-\ln\left(1+\frac p{\sigma^2}\right)
\right]\log e,
\qquad p\geq0.
\]
Then \(f_n(0)=0\), \(f_n(p)\to\infty\) as
\(p\to\infty\), and
\[
f_n'(p)
=
\frac{np}{2\sigma^2(\sigma^2+p)}\log e
>0
\qquad (p>0).
\]
Thus \(f_n(p)=\delta\) has a unique positive
solution \(\Psi(\delta,\sigma,n)\).
The covertness constraint implies
\[
f_n(\overline P_{\bm c})\leq\delta,
\]
and consequently
\[
\overline P_{\bm c}
\leq\Psi(\delta,\sigma,n).
\]
\end{proof}
\subsection{Proof of Corollary 2}
\begin{proof}
Put
\[
N:=MK,\qquad
p_n:=\Psi(\delta,\sigma,n),\qquad
t_n:=\frac{n}{n-1}p_n.
\]
Define the low-energy index set
\[
\mathcal D:=
\left\{
(s,w):
\frac{\|\bm x_{sw}\|^2}{n}\leq t_n
\right\}.
\]
Suppose, for contradiction, that
\[
|\mathcal D|<
\left\lceil\frac Nn\right\rceil.
\]
Since \(|\mathcal D|\) is an integer,
\[
|\mathcal D|
\leq\left\lceil\frac Nn\right\rceil-1.
\]
Therefore,
\[
|\mathcal D^c|
\geq
N-\left\lceil\frac Nn\right\rceil+1
>
N\left(1-\frac1n\right).
\]
Every index in \(\mathcal D^c\) has energy
per channel use strictly larger than \(t_n\).
Consequently,
\begin{align*}
\overline P_{\bm c}
&=
\frac1N
\sum_{s,w}\frac{\|\bm x_{sw}\|^2}{n}\\
&\geq
\frac1N
\sum_{(s,w)\in\mathcal D^c}
\frac{\|\bm x_{sw}\|^2}{n}\\
&>
\frac{|\mathcal D^c|}{N}t_n\\
&>
\left(1-\frac1n\right)
\frac{n}{n-1}p_n\\
&=p_n.
\end{align*}
This contradicts the converse average-power
lemma. Hence
\[
|\mathcal D|
\geq\left\lceil\frac{MK}{n}\right\rceil.
\]
The energy bound follows directly from
the definition of \(\mathcal D\).
\end{proof}
\subsection{Proof of Theorem 4}
\begin{proof}
Fix $\delta>0$, $\sigma>0$, and $0<\epsilon<1$.
Consider any $(M,K,n,\epsilon,\sigma,\delta)$ code.
By Lemma 14 and
Corollary 2, there is an index set
\(\mathcal D\) satisfying
\[
|\mathcal D|\geq
\left\lceil\frac{MK}{n}\right\rceil
\]
and
\[
\frac{\|\bm x_{sw}\|^2}{n}\leq p_n
\quad\text{for every }(s,w)\in\mathcal D.
\]
For each key \(s\), define
\[
\mathcal D_s:=
\{w:(s,w)\in\mathcal D\}.
\]
Since
\[
\sum_{s=1}^{K}|\mathcal D_s|
=|\mathcal D|,
\]
there exists a key \(s_*\) such that
\begin{equation}\label{upper111}
|\mathcal D_{s_*}|
\geq\frac{|\mathcal D|}{K}
\geq\frac Mn.
\end{equation}
Fix this key and retain only the messages
in \(\mathcal D_{s_*}\).
Use its original decoder, treating any output
outside \(\mathcal D_{s_*}\) as an error.
For every retained message, its probability
of correct decoding is unchanged.
Thus the retained code has maximal decoding
error probability at most \(\epsilon\).
Every retained codeword also satisfies $\|\bm x_{s_*w}\|^2\leq np_n$.
The conventional converse bound of finite block AWGN code
under a maximal-power constraint can not be directly applied, since $p_n$ is a varying function of $n$. We have
\[
|\mathcal D_{s_*}|
\leq M_m^*(n,\epsilon,p_n).
\]
Combining this with~\eqref{upper111} gives
\begin{equation}\label{eq:converse-subcode-loss}
\log M
\leq
\log M_m^*(n,\epsilon,p_n)+\log n.
\end{equation}
Applying the converse with varying maximal-power 
\cite[Eq.~(24)]{Yu1}:
\begin{align}
\log M_m^*(n,\epsilon,P(n))
\leq\frac n2\log(1+P(n)) \\
+\sqrt{n\mathsf V(P(n))}\Phi^{-1}(\epsilon)+ \frac{1}{2}\log n + O(1), \nonumber
\end{align}
with $P(n) \rightarrow 0$ as $n\rightarrow \infty$, the following inequality 
\begin{align}\label{converse2}
\log|\mathcal D_{s_*}|
\leq{}&
\frac{n}{2}\log(1+p_n) \\
&+\sqrt{n\mathsf V(p_n)}\,\Phi^{-1}(\epsilon)
+ \frac{1}{2}\log n + O(1), \nonumber
\end{align}
with 
$\mathsf V(P)
=\frac{P(P+2)}{2(1+P)^2}(\log e)^2$ 
holds for $p_n = O(n^{-\frac{1}{2}})$.

Here, we have used
$-\Phi^{-1}(1-\epsilon)=\Phi^{-1}(\epsilon)$.
We next expand the power bound.
Let $a=\delta\ln 2$ and
$x_n=\Psi(\delta,\sigma,n)/\sigma^2$.
By (\ref{averagepowerlevel}) of the main text, $x_n$ is the unique positive
solution of
\begin{equation}\label{eq:power_equation}
\frac{n}{2}\big[x_n-\ln(1+x_n)\big]\log e=\delta.
\end{equation}
Equivalently,
\[
x_n-\ln(1+x_n)=\frac{2a}{n}.
\]
Using
\[
x-\ln(1+x)
=\frac{x^2}{2}-\frac{x^3}{3}+O(x^4),
\]
and matching the coefficients of $n^{-1}$ and
$n^{-3/2}$ yields
\begin{equation}
x_n=2\sqrt a\,n^{-1/2}
+\frac{4a}{3}n^{-1}
+O(n^{-3/2}).
\end{equation}
Consequently, we arrive at,
\begin{equation}\label{eq:power_expansion}
p_n
=2\sigma^2\sqrt a\,n^{-1/2}
+\frac{4}{3}\sigma^2a\,n^{-1}
+O(n^{-3/2})
\end{equation}
by using  $n/(n-1)=1+O(n^{-1})$. Therefore, we have $p_n = O(n^{-\frac{1}{2}})$. Taylor's theorem now gives
\begin{align}\label{eq:capacity_expansion}
\frac{n}{2}\log(1+p_n)
&=\frac{n}{2}
\left(p_n-\frac{p_n^2}{2}+O(p_n^3)\right)\log e
\nonumber\\
&=\sigma^2\sqrt{an}\log e
+\left(\frac{2}{3}\sigma^2-\sigma^4\right)a\log e \nonumber\\
 & + O(n^{-1/2})
\nonumber\\
&=\sigma^2(n\delta\log e)^{1/2}+O(1).
\end{align}
Moreover, we have
$\mathsf V(P)
=\big(P+O(P^2)\big)(\log e)^2 \, \  \,\text{as }P\to0$.
Together with \eqref{eq:power_expansion}, this implies
\[
n\mathsf V(p_n)
=\left(2\sigma^2\sqrt a\,\sqrt n+O(1)\right)
(\log e)^2,
\]
and therefore
\begin{align}\label{eq:dispersion_expansion}
\sqrt{n\mathsf V(p_n)}
&=\sqrt{2}\,\sigma a^{1/4}n^{1/4}\log e
+O(n^{-1/4})
\nonumber\\
&=\sqrt{2}\,\sigma(n\delta)^{1/4}
(\log e)^{3/4}+O(n^{-1/4}).
\end{align}
Substituting \eqref{eq:capacity_expansion} and
\eqref{eq:dispersion_expansion} into \eqref{converse2}
yields
\begin{align}
\log&|\mathcal D_{s_*}|
\leq 
\sigma^2(n\delta\log e)^{1/2}
\nonumber\\
&+\sqrt{2}\,\sigma(n\delta)^{1/4}
(\log e)^{3/4}\Phi^{-1}(\epsilon)
+O(\log n).
\end{align}
Finally, by \eqref{upper111}, we get
\[
\log M
\leq\log|\mathcal D_{s_*}|+\log n.
\]
Since the original code was arbitrary, absorbing
$\log n$ into the remainder establishes
\begin{align}
\log & M^*(n,\epsilon,\sigma,\delta)
\leq
\sigma^2(n\delta\log e)^{1/2}
\nonumber\\
&+\sqrt{2}\,\sigma(n\delta)^{1/4}
(\log e)^{3/4}\Phi^{-1}(\epsilon)
+O(\log n).
\end{align}
 \end{proof}

\subsection{Decay Rates of the Truncation Penalties with constant $\mu$}
In this part, assume $\sigma^2>0$, $\Psi(n)>0$, and fixed $\mu\in(1/2,1)$. We prove that the normalization probability $\Delta$ tends to $1$ exponentially. As a consequence, we show all the quantities $1-\Delta$, $-\log\Delta$, and $r_n-\log\Delta$ tend to zero exponentially.
\subsubsection{Normalization probability}
From the definition, 
$\mathcal{A}_n=\{\bm x:\mu^2n\Psi(n)\le\|\bm x\|^2\le n\Psi(n)\}$ and
$\Delta$ is the probability content of $\mathcal{A}_n$ under $P_n^{(n)}$,
we have $\Delta=P_n^{(n)}(\mathcal{A}_n)$.

Under $P_n^{(n)}$, It is obvious that $U_n \triangleq \|\bm X\|^2/(\mu\Psi(n))\sim\chi_n^2$. Thus
\begin{align}
 \Delta=\mathbb{P}(\mu n\le U_n\le n/\mu),\nonumber \\
 \varepsilon_n\triangleq 1-\Delta = \mathbb{P}(U_n<\mu n)+ \mathbb{P}(U_n>n/\mu).
\end{align}
In particular, $\Delta$ depends only on $n$ and $\mu$, since the positive power scale cancels.
\subsubsection{Chernoff bound on $1-\Delta$}
The moment-generating function of $U_n$ is $\mathbb{E} e^{tU_n}=(1-2t)^{-n/2}$ for $t<1/2$. By Chernoff bounding, for $s>0$,
\[
 \mathbb{P}(U_n\le\mu n)\le e^{s\mu n}(1+2s)^{-n/2}.
\]
The optimizer is $s=(1-\mu)/(2\mu)$, giving
\begin{equation}
 \mathbb{P}(U_n\le\mu n)\le e^{-nc_-(\mu)},\qquad
 c_-(\mu)=\frac{\mu-1-\ln\mu}{2}.
\end{equation}
For $0<t<1/2$,
\[
 \mathbb{P}(U_n\ge n/\mu)\le e^{-tn/\mu}(1-2t)^{-n/2}.
\]
The optimizer is $t=(1-\mu)/2$, giving
\begin{equation}
\mathbb{P}(U_n\ge n/\mu)\le e^{-nc_+(\mu)},\, \  \,
 c_+(\mu)=\frac{\mu^{-1}-1+\ln\mu}{2}.
\end{equation}
As 
\begin{equation}
c_-(\mu)=\frac12\int_\mu^1\left(\frac1t-1\right)\,dt>0,
 \qquad 0<\mu<1.
\end{equation}
 More generally, the function $v-1-\ln v$ is strictly positive for all $v>0$ with $v\ne1$: its derivative is $1-1/v$, so it has its unique minimum, equal to zero, at $v=1$. Applying this observation to $v=1/\mu$ also proves $c_+(\mu)>0$ for $0<\mu<1$.

Furthermore, $c_+(\mu)>c_-(\mu)$ for $0<\mu<1$: the difference is $(\mu^{-1}-\mu+2\ln\mu)/2$, which vanishes at one and has derivative $-(1-\mu)^2/(2\mu^2)<0$. Hence, we get
\begin{equation}\label{eq:tail111}
\varepsilon_n\le e^{-nc_-(\mu)}+e^{-nc_+(\mu)}\le2e^{-nc_-(\mu)}.
\end{equation}
\subsubsection{On the bound of $-\log\Delta$}
For $0 \le u < 1$, the inequality $u \le -\ln(1-u) \le \frac{u}{1-u}$ holds. 
Substituting $u = \varepsilon_n$ and noting that $\Delta = 1 - \varepsilon_n$, we obtain
\begin{equation}\label{eq:log}
  \varepsilon_n \log e\le -\log\Delta \le \,\frac{\varepsilon_n}{1-\varepsilon_n} \log e.
\end{equation}
For all sufficiently large $n$, $\varepsilon_n\le1/2$, so
\begin{equation}\label{eqlog222}
 -\log\Delta\le2\varepsilon_n\log e\le 4e^{-nc_-(\mu)} \log e.
\end{equation}
\subsubsection{On the bound of $r_n-\log\Delta$}\label{boundonr_n-logdelta}
Define $L_n=\frac{n\mu^2\Psi(n)^2}{2\sigma^2[\sigma^2+\mu\Psi(n)]}\log e$, and then $r_n=L_n\varepsilon_n/(1-\varepsilon_n)$. If $\Psi(n)\le C/\sqrt n$ for some constant $C$ eventually, then
\[
 0\le L_n\le L\triangleq\frac{\mu^2C^2}{2\sigma^4}\log e.
\]
Consequently, we get $r_n\le4L e^{-nc_-(\mu)}$ from (\ref{eq:tail111}). Combining this with \eqref{eq:log} yields the bound
\begin{equation}
0\le r_n-\log\Delta\le(L_n+\log e)\frac{1-\Delta}{\Delta}.
\end{equation}
For all sufficiently large $n$, by using (\ref{eqlog222}) we arrive at
\begin{equation}\label{eq:combined}
r_n-\log\Delta\le4(L+\log e)e^{-nc_-(\mu)}.
\end{equation}
In addition, $L$ and $c_-(\mu)>0$ are independent of $n$ for fixed $\mu$.

\end{document}